\RequirePackage{amsmath}
\documentclass[12pt,a4paper]{iopart}
\usepackage{graphicx}
\usepackage{cite}
\usepackage{color}
\usepackage{multirow}
\usepackage{epsfig}
\usepackage{epstopdf}
\usepackage{iopams}
\usepackage{cancel}
\usepackage{multicol}
\usepackage{ipa}
\usepackage{mathabx}

\usepackage[breaklinks=true,colorlinks=true,linkcolor=blue,urlcolor=blue,citecolor=blue]{hyperref}

\usepackage[deletedmarkup=xout]{changes}
\definechangesauthor[color=red]{RPB}
\definechangesauthor[color=cyan]{SW}
\definechangesauthor[color=blue]{jTr}
\definechangesauthor[color=magenta]{TM}
\definechangesauthor[color=orange]{JJ}

\begin{document}

\title[Ion dynamics in Ar/Xe CCPs]{Ion dynamics in capacitively coupled argon-xenon discharges}

\author{M. Klich$^{1}$, S. Wilczek$^{1}$,  J. F. J. Janssen$^{2}$, R. P. Brinkmann$^{1}$, T. Mussenbrock$^{1}$, J. Trieschmann$^{3}$}

\address{$^1$Department of Electrical Engineering and Information Science, Ruhr University Bochum, D-44780, Bochum, Germany}
\address{$^2$PlasmaMatters B.V., De Groene Loper 19, 5600 MB Eindhoven, The Netherlands}
\address{$^3$Electrodynamics and Physical Electronics Group, Brandenburg University of Technology Cottbus-Senftenberg, Siemens-Halske-Ring 14, 03046 Cottbus, Germany}

\date{\today}

\begin{abstract}
	An argon-xenon (Ar/Xe) plasma is used as a model system for complex plasmas.
	Based on this system, symmetric low-\-pressure capacitively coupled ra\-dio\-frequency discharges are examined utilizing Particle-In-Cell/Monte Carlo Collisions (PIC/MCC) simulations.
	In addition to the simulation, an analytical energy balance model fed with the simulation data is applied to analyze the findings further.
	This work focuses on investigating the ion dynamics in a plasma with two ion species and a gas mixture as background.
	By varying the gas composition and driving voltage of the single-frequency discharge, fundamental mechanics of the discharge, such as the evolution of the plasma density and the energy dispersion, are discussed.
	Thereby, close attention is paid to these measures' influence on the ion energy distribution functions at the electrode surfaces.
	The results show that both the gas composition and the driving voltage can significantly impact the ion dynamics.
	The mixing ratio of argon to xenon allows for shifting the distribution function for one ion species from collisionless to collision dominated.
	The mixing ratio serves as a control parameter for the ion flux and the impingement energy of ions at the surfaces.
	Additionally, a synergy effect between the ionization of argon and the ionization of xenon is found and discussed.
\end{abstract}

%%%%%%%
\section{Introduction}

Radio-frequency capacitively coupled plasmas (RF-CCPs) operated at low-pressures are a core part of modern technology \cite{LiebermanBook, ChabertBook, MakabeBook}.
Especially for semiconductor fabrication, plasma processes like ion-assisted etching \cite{HandbookEtching, etching1} and ion implantation \cite{implantation1, implantation2, implantation3} are key technologies.
Plasma tools help to achieve an integration depth of only a few nanometers \cite{dev-scale1, dev-scale2, dev-scale3}.
One major challenge of these processes is to control the energy and flux of impinging ions on the wafer separately \cite{LiebermanBook, ChabertBook, MakabeBook, IEDFcontrol1, IEDFcontrol2, IEDFcontrol3, IEDFcontrol4, IEDFcontrol5}.\par
Techniques using multiple driving frequencies, such as voltage waveform tailoring \cite{IEDFcontrol1}, succeed to independently control the plasma generation and the ion bombardment energy \cite{MF1, MF2}.
The plasmas investigated in these studies are predominantly argon plasmas \cite{IEDFcontrol2, IEDFcontrol5, VWT1}.
However, industrially relevant etching plasmas consist of rather complex gas mixtures like CF\textsubscript{4}/H\textsubscript{2} \cite{chemistry1, chemistry2, chemistry3} or SF\textsubscript{6}/O\textsubscript{2} \cite{chemistry4, chemistry5}.
For these plasmas, the interplay of several charged and neutral heavy species impacts the ion dynamics.
The ion dynamics in the plasma eventually determine how ions reach the walls.
Here, both the quantitative (e.g., how many ions reach the target/substrate?) and the qualitative perspective (e.g., how are the ions affected by collisions?) need to be considered.\par
Researching complex plasma chemistry in RF-CCPs is a tedious task.
Experimental studies show that the ion energy distribution functions (IEDFs) at the electrodes become rather complicated \cite{IEDF1, IEDF2, IEDF3, Kawamura, IEDF5}.
Commonly used tools such as the retarding field analyzer filter the incident ions by energy and do not differentiate between the ion species \cite{IEDF1, IEDF5, IEDF6}.
There is recent and ongoing work to utilize ion mass spectrometry to overcome these issues \cite{IEDF7}.
Nevertheless, this technique is currently not widely applied as a diagnostic tool to analyze plasmas.
Therefore, theoretical studies and simulations are necessary to help to interpret and to understand the measured data.
However, the inherently complex chemistry renders a complete simulation cumbersome.
The commonly used kinetic Particle-In-Cell/Monte Carlo collisions (PIC/MCC) method typically avoids complex chemistry mainly due to lack of cross section data (although conceptually possible).
The reasoning is to keep the number of species and superparticles traceable \cite{Kim}.
Otherwise, the computational load of PIC/MCC simulation would not be feasible.\par
Combining complex discharge chemistry with the multi-frequency approaches mentioned above makes a detailed assessment of ion dynamics' features too cumbersome to conduct collectively.
Hence, we decide to investigate the fundamental principles of a discharge with two ion species for this study.
The mixture of the noble gases argon and xenon has some history of being an adequate model for complex chemistry.
In low-pressure plasmas, the plasma chemistry of noble gases becomes relatively simple \cite{Gudmundsson1, Gudmundsson2}.
Therefore, studies on ion acoustic waves \cite{IAW1, IAW2} and a generalized Bohm criterion \cite{Bohm1, Gudmundsson1, Gudmundsson2} depicted this mixture as a simple example for a multi-ion discharge.
Recent studies by Kim et al.\cite{Kim2} and Adrian et al.\cite{Adrian} contributed to those discussions using or referring to Ar/Xe plasmas.\par
Apart from being a model system, there are some academic applications of Ar/Xe plasmas (e.g., as trace gas for mass spectrometry \cite{PlasmaTorch, MassSpec}, for the diagnostics of the electron temperature \cite{ ElTemp}, or in halide lamp simulations \cite{Lamp}).
Furthermore, the mixture has had great success as the illuminant \cite{ArXePDP} or as part of the illuminant mixture \cite{Kim, PDP1, PDP2} of plasma display panels (PDPs).
This historical background causes both gases to be relatively well researched.
This fact entails many valuable data for theory and simulation.\par
This work aims to add to the existing studies conducted for various gas mixtures \cite{mixture1, mixture2, mixture3, mixture4}, investigating their intrinsic mixture dynamics.
This knowledge will eventually enable the adaptation of the known means of plasma control to the complex discharges of industrial relevance.
In contrast to the existing studies, our work is focused on the impact of the gas mixture composition on the ion dynamics.
We will show that the gas composition is a suitable control parameter for the ion dynamics (e.g., the impingement energy of ions at the surface).\par
This manuscript presents our findings as follows: in section \ref{methods}, we introduce our simulation framework and a model for the missing cross section data.
Moreover, we introduce an energy balance model for CCPs with multiple ion species.
The findings of these models are interpreted in section \ref{results}.
We first discuss the influence of a variation of the gas composition on the ion dynamics.
Then, we validate the energy balance model with our simulation data.
Afterward, we apply this energy balance model to support and to analyze our gas composition variation findings.
We conclude section \ref{results} by discussing and examining the influence of a variation of the gas composition combined with a variation of the driving voltage on the ion dynamics.
Finally, in section \ref{conclusion}, we summarise our findings, draw a conclusion, and set this work into the context of industrial applications.

%%%%%%%
\section{Methods and models} \label{methods}

\subsection{Particle-In-Cell Simulation} \label{PIC_sec}
	The first particle simulations were introduced in the 1940s \cite{HockneyBook}, and the PIC/MCC scheme was developed in the 1960s \cite{Birdsall}.
	Since then, the PIC/MCC method became a commonly used tool to self-consistently simulate low-pressure plasmas \cite{Kim, Birdsall, Verboncoeur}.
	Despite having the disadvantage of a substantial computational load, its most significant advantage is the statistical representation of distribution functions in phase-space, allowing the method to capture non-local dynamics \cite{Verboncoeur, Wilczek1}.\par
	For this work, a benchmarked PIC/MCC implementation called \textit{yapic1D} \cite{PICbenchmark} is used to generate the results.
	The original code is modified to include two background gases and multiple ion species.
	Aside from that, diagnostics for the energy balance model mentioned above is added to the original code.\par
	This simulation setup is taken to be fully geometrically symmetric (compare Wilczek et al. for details \cite{Wilczek1}).
	1d3v electrostatic simulations are executed using a Cartesian grid with 800 grid cells representing an electrode gap of $25\,$mm.
	The resulting cell size $\Delta{x}$ meets the requirement to resolve the Debye length $\lambda_\mathrm{D}$ \cite{Birdsall, PICbenchmark, Wilczek1}.
	Similarly, the single harmonic driving frequency $f_\mathrm{RF} = 13.56\,$MHz is sampled with 3000 points per RF period.
	The time step $\Delta{t}$ is sufficiently small to fulfill the requirement regarding the electron plasma frequency $\omega_\mathrm{pe}$ \cite{Birdsall, PICbenchmark, Wilczek1}.
	Several other studies mention the influence of the number of superparticles on the statistics and the plasma density \cite{PICbenchmark, Erden, Kim-PIC}.
	For this work, we did not include individual weighting for different particle species.
	To have an acceptable resolution for each ion species at all values of the xenon fraction $x_\mathrm{Xe}$, we simulated about 800.000 super-electrons for each case.
	The advantage of this choice is that an average of 3000 converged RF-cycles provides satisfactory results.\par
	The ideal gas law defines the neutral species' total density, and the neutral fraction $x_i$ is varied.
	Thereby, the gas pressure $p_\mathrm{gas}$  is kept constant at $3\,$Pa, and the gas temperature $T_\mathrm{gas}$ at $300\,$K.
	First, we choose the amplitude of the RF voltage $V_\mathrm{RF}$ to be $100\,$V.
	Later in section \ref{VolVar}, we discuss the implications of a voltage variation between $100\,$V and $1000\,$V on the ion dynamics.
	All the parameters presented in this section are typical for baseline studies of RF-CCPs \cite{IEDFcontrol3, IEDFcontrol4, IEDFcontrol5, PICbenchmark, Wilczek1}.
	
%%%%%%%

\subsection{Discharge chemistry} \label{chemistry}
    
    \begin{table}[t!]
        \centering
        \begin{tabular}{l l l c c}\hline
            $\hash$ & reaction & process name & $\varepsilon_\mathrm{thr}$ [eV] & data source \\\hline
            1 & e$^-$ + Ar $\rightarrow$ e$^-$ + Ar & elastic scattering & - & Phelps \\
            2 & e$^-$ + Ar $\rightarrow$ e$^-$ + Ar$^*$ & electronic excitation & 11.5 & Phelps \\
            3 & e$^-$ + Ar $\rightarrow$ 2$\,$e$^-$ + Ar$^+$ & ionization & 15.8 & Phelps \\
            4 & e$^-$ + Xe $\rightarrow$ e$^-$ + Xe & elastic scattering & - & Phelps \\
            5 & e$^-$ + Xe $\rightarrow$ e$^-$ + Xe$^*$ & electronic excitation & 8.32 & Phelps \\
            6 & e$^-$ + Xe $\rightarrow$ 2$\,$e$^-$ + Xe$^+$ & ionization & 12.12 & Phelps \\\hline
            7 & Ar$^+$ + Ar $\rightarrow$ Ar$^+$ + Ar & isotropic scattering & - & Phelps \\
            8 & Ar$^+$ + Ar $\rightarrow$ Ar + Ar$^+$ & resonant charge exchange & - & Phelps \\
            9 & Ar$^+$ + Xe $\rightarrow$ Ar$^+$ + Xe & isotropic scattering & - & LJ pot \\
            10 & Xe$^+$ + Xe $\rightarrow$ Xe$^+$ + Xe & isotropic scattering & - & Phelps \\
            11 & Xe$^+$ + Xe $\rightarrow$ Xe + Xe$^+$ & resonant charge exchange & - & Phelps \\
            12 & Xe$^+$ + Ar $\rightarrow$ Xe$^+$ + Ar & isotropic scattering & - & Viehland\\\hline
        \end{tabular}
	    \caption{Plasma chemistry and collision processes considered in the simulation.
	    	Meaning of the data sources: ``Phelps'' refers to the cross section data found initially in the JILA database \cite{Phelps} and now distributed by the LXCat project \cite{lxcat1, lxcat2, lxcat3}.
		``LJ pot'' refers to a cross section obtained based on a phenomenological Lennard-Jones potential as described by Laricchiuta et al.\cite{Laricchiuta}.
		``Viehland'' marks a cross section calculated from an interaction potential given by Viehland et al.\cite{Viehland}.
		Details of the calculations can be found in section \ref{xSectData}.}
        \label{tab:chemistry}
    \end{table}

     For PIC simulations to provide a realistic representation of the particle distribution functions and physics in a low-pressure discharge, collisions need to be considered.
     The method of choice is the Monte Carlo collision technique \cite{Birdsall, Verboncoeur, Wilczek1} that is combined with a so-called null collision scheme \cite{Skullerud, PICbenchmark, Wilczek1}.
     Both techniques require the knowledge of momentum transfer cross sections.\par
	The chemistry set for argon and xenon is in line with the work of Gu\eth mundsson et al.\cite{Gudmundsson1, Gudmundsson2}.
	All reactions can be seen in detail in table \ref{tab:chemistry}.
	In contrast to Gu\eth mundsson et al., we decide to take advantage of the commonly used and acknowledged \cite{Wilczek1, PICbenchmark, VWT1, DonkoPIC} cross section data obtained by Phelps.
	The data was initially distributed via the JILA database \cite{Phelps} and is now available at the LXCat project website \cite{lxcat1, lxcat2, lxcat3}.
	Phelps combines the cross sections for all electronically excited states into one ``effective excitation'' cross section.
	This effective excitation reduces the total number of reactions and the numerical load.\par
	The second difference compared to Gu\eth mundsson et al. is our treatment of the missing cross section data for Ar$^+$/Xe and Xe$^+$/Ar.
	Both conclude to neglect charge transfer collisions between argon and xenon due to it being a non-resonant process that requires a third particle to ensure momentum and energy conservation.
	The disparity lies in our treatment of the remaining scattering process.
	They assume the cross sections for processes 7 and 9 and the cross sections for processes 10 and 12, respectively, to be equal.
	We adopt a physical model to procure the necessary cross sections from interaction potentials.
	In this way, we create individual cross sections for processes 9 and 12.
	The details of how to calculate these cross sections will be presented in the following section. \par
\begin{figure*}[t!]
    	\begin{center}
           	\includegraphics[width = \textwidth]{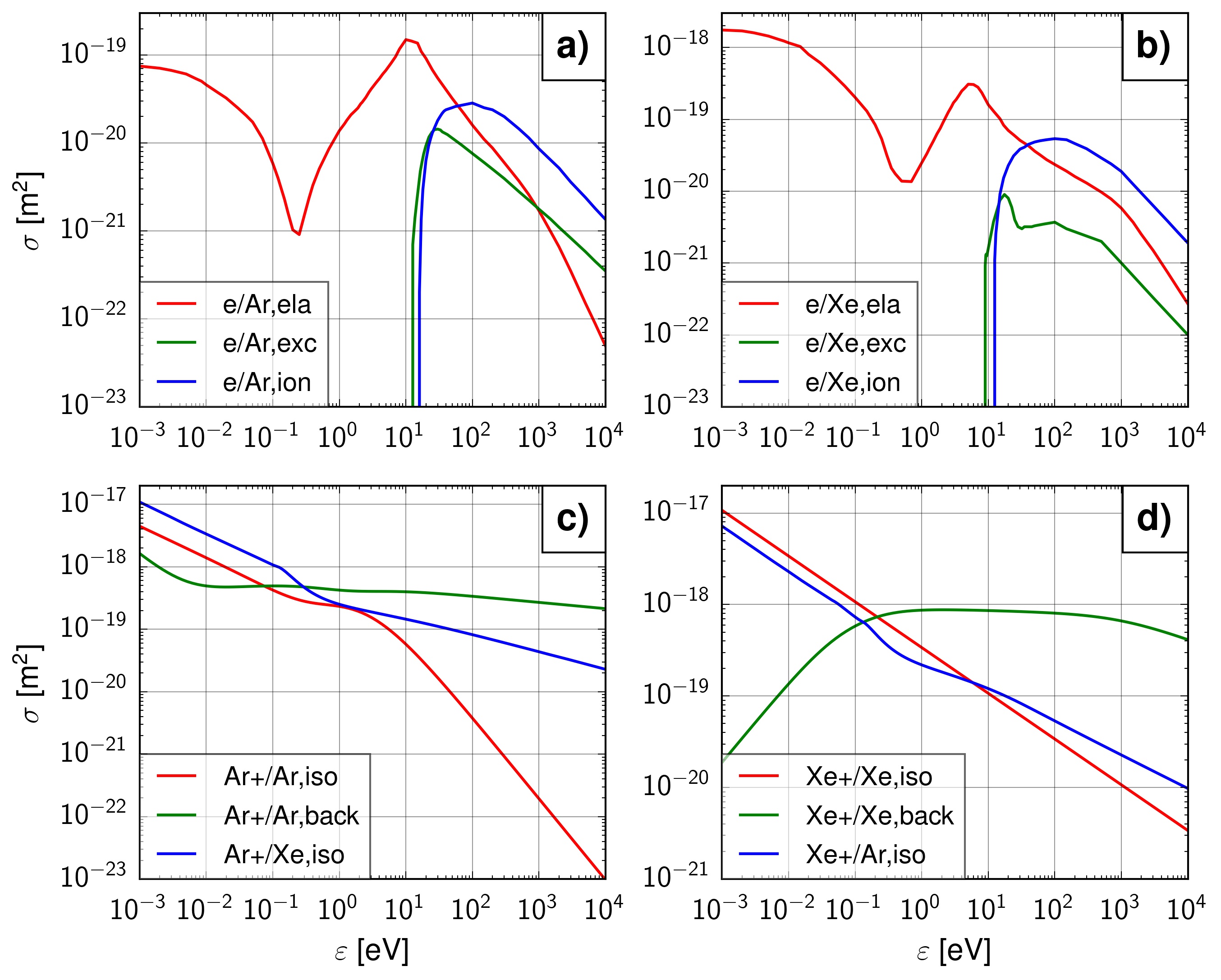}
    		\caption{Cross sections for the electron and ion collisions used in this work.
			a) shows the cross section of collision processes from electrons with argon neutrals.
			b) shows the cross sections of collision processes from electrons with xenon neutrals.
			c) shows cross sections of the collisions of Ar\textsuperscript{+} ions.
			d) shows cross sections of the collisions of Xe\textsuperscript{+} ions.
			The data source and a detailed description of each process are found in table \ref{tab:chemistry}.
			Abbreviations used in the legend:
				\textit{ela} = elastic collision electron/neutral,
				\textit{exc} = electronic excitation electron/neutral,
				\textit{ion} = electron impact ionization,
				\text{iso} = isotropic scattering ion/neutral as defined by \cite{Phelps},
				\textit{back} = backscattering ion/neutral as defined by \cite{Phelps}.}
    		\label{fig:crosssections}
    	\end{center}
    \end{figure*}

	The cross section data used for this work are depicted in figure \ref{fig:crosssections}.
	It is noticeable that the cross sections of processes involving xenon species generally have higher values than corresponding processes that involve argon species.
	In terms of a hard-sphere model \cite{LiebermanBook}, this deviation is explained by the different covalent atom radii of argon and xenon \cite{CRCHandbook}.
	Xenon is, compared to argon, simply the bigger target.
	In terms of a more sophisticated collision model \cite{LiebermanBook}, one, for example, needs to consider the atomic polarisability of the neutral particle.
	Nevertheless, such a view leads to the same insight.
	Xenon has a higher atomic polarisability than argon \cite{CRCHandbook} and stronger interaction with charged particles.
	Correspondingly, the cross section for charged particles interacting with xenon has to be larger than the cross section for the interaction of charged  particles and argon.
	
%%%%%%%%%%%%%%%	
    
\subsection{The calculation of the cross sections} \label{xSectData}
    On an elementary level of theory, all cross sections are based on an interaction potential between the colliding particles.
    If the literature does not provide a cross section, a possible solution is to make it the modeler's task to develop an interaction potential by making several assumptions.
    A classic example of this is the Langevin capture cross section \cite{Langevin} used in studies to make up for unknown cross sections \cite{Donko2}.
    Despite the Langevin cross section's advantages, a complete implementation is numerically extensive and leads to anisotropic scattering \cite{Nanbu1, Nanbu2}.
    The cross sections given by Phelps are a kind of momentum transfer cross sections \cite{Phelps}.
    There, the scattering angles are found in an isotropic manner \cite{Birdsall}.
    Hence, it is questionable to apply anisotropic scattering for two collisions while the other collisions are treated isotropically.
    We perceive another approach to be more suitable for this work.\par
    The approach used in this work is based on Laricchiuta et al. \cite{Laricchiuta}, who use a phenomenological potential to describe a two-body interaction given by
	\begin{align}
		V_{ij}(x) = \epsilon_{p,ij}\left[\frac{m}{n_{ij}\left(x_{ij}\right)-m}\left(\frac{1}{x_{ij}}\right)^{n_{ij}\left(x_{ij}\right)}-\frac{n_{ij}\left(x_{ij}\right)}{n_{ij}\left(x_{ij}\right)-m}\left(\frac{1}{x_{ij}}\right)^{m}\right],
	\end{align}
    where the standard exponents of the Lennard Jones potential, 12 and 6, are replaced by $n(x_{ij})$ and $m$.
    Depending on the type of interaction, $m$ is either 4 for neutral-ion interactions or 6 for neutral-neutral interactions.
    In this work, the potential is applied to neutral-ion interactions only.
    Hence, m is always equal to 4.
    The dimensionless coordinate $x=r/r_{m,ij}$ depends on the parameterized position of the potential well $r_{p,ij}$.
    The potential itself is scaled by the parameterized potential well depth $\epsilon_{p,ij}$.
    Both parameterizations are empirical approximations that depend on atomic properties like the polarizability.
    More details related to the exact empirical formulas can be found in Laricchiuta et al. \cite{Laricchiuta}, Cambi et al. \cite{Cambi}, Cappelletti et al. \cite{Cappelletti}, and Aquilanti et al. \cite{Aquilanti}.\par
    Two additional steps are required to obtain the cross section.
    The first step is calculating the scattering angle, $\chi_{ij}$ according to
	\begin{align}
		\chi_{ij}\left(\epsilon_{ij},b\right) = \pi -2 b \int_{r_0}^\infty
		\frac{\text{d}r}{r^2
		\sqrt{1-\frac{b^2}{r^2}-\frac{V\left(r\right)}{\epsilon_{ij}}}}
	\end{align}
    with $\epsilon_{ij}$ the kinetic energy in the center of mass frame, $b$ the impact parameter, $r$ the distance between the particles, and $r_0$ the distance of closest approach.
    The scattering angles are calculated using a program that is based on Colonna et al. \cite{Colonna}.
    The second step is calculating the cross section $\sigma_{ij}$
	\begin{align}
		\sigma_{ij}^{\left(l\right)}\left(\epsilon_{ij}\right)=2 \pi \int_0^\infty
		\left[1-\cos^l \chi_{ij}\left(\epsilon_{ij},b\right)\right]b\text{d}b,
	\end{align}
    with $l$ an integer that indicates which type of cross section is calculated.
    In this work, we used $l=1$, which corresponds to the momentum transfer cross section.
    The cross section is integrated based on an algorithm developed by Viehland \cite{Viehland2010}.
    Finally, the scattering angle corresponding to the obtained momentum transfer cross sections is consistently taken to be isotropic in our simulations.
    
%%%%%%%%%%%%%%%
    
\subsection{Energy balance model} \label{EB_sec}
    The conservation of energy is one of the central continuity equations of physics and so knowing how the energy disperses into a system is key to understanding the process.
    In terms of low-temperature plasma physics, a frequently used model as given by Lieberman and Lichtenberg \cite{LiebermanBook} for a geometrically symmetric situation reads:
    \begin{align}
        S_\mathrm{abs} = 2\,n_\mathrm{s}\,u_\mathrm{B}\,\varepsilon_\mathrm{tot} = 2\,\Gamma_\mathrm{B}\,\varepsilon_\mathrm{tot} = 2\,e\,\Gamma_\mathrm{B}\,( \varepsilon_\mathrm{e} + \varepsilon_\mathrm{c} + \varepsilon_\mathrm{i} ).
        \label{eq:cl_energybalance}
    \end{align}
    $S_\mathrm{abs}$ denotes the total energy flux into the system,
    $n_\mathrm{s}$ the plasma density at the sheath edge,
    $u_\mathrm{B}$ denotes the Bohm velocity,
    $\Gamma_\mathrm{B}$ is the ion flux at the Bohm point,
    and $\varepsilon_\mathrm{tot}$ is the total energy loss in eV.
    The last transformation in equation \eqref{eq:cl_energybalance} shows that the energy loss per electron-ion pair created may be split into
    an energy loss due to electrons hitting the bounding surface ($\varepsilon_\mathrm{e}$),
    an energy loss due to collisions ($\varepsilon_\mathrm{c}$), and
    an energy loss due to ions impinging at the bounding surface ($\varepsilon_\mathrm{i}$).
    The loss terms $\varepsilon_\mathrm{e}$ and $\varepsilon_\mathrm{i}$ describe an averaged energy loss of the system per lost particle (neglecting particle reflections).
    The third term $\varepsilon_\mathrm{c}$ is treated differently.
    It represents the collisional losses per newly created electron/ion pair.\par
    Previous work \cite{Wilczek2} has shown that an adaptation of equation \eqref{eq:cl_energybalance} gives insight into the system's electron dynamics by calculating all necessary terms from a PIC/MCC simulation.
    An essential insight is that, due to flux conservation, the Bohm flux $\Gamma_\mathrm{B}$ can be exchanged by the electron flux $\Gamma_\mathrm{e,el}$ or the ion flux $\Gamma_\mathrm{i,el}$ at the electrode.\par
    In detail, the energy conversion through collisions $\varepsilon_\mathrm{c}$ consists of an electron $\varepsilon_\mathrm{c,e}$ and an ion contribution $\varepsilon_\mathrm{c,i}$.
    For low-pressure plasmas, it is argued that the energy loss due to ion collisions $\varepsilon_\mathrm{c,i}$ is often negligible \cite{LiebermanBook}.
    However, a PIC/MCC study by Jiang et al.\cite{Jiang} showed that $\varepsilon_\mathrm{c,i}$ can significantly impact the energy balance of low-pressure plasmas.\par
    Using both insights, we evolve equation \eqref{eq:cl_energybalance} into an energy balance model for two ion species, here explicitly given for our case of an Ar/Xe mixture:
    \begin{alignat}{3}
        &S_\mathrm{abs,tot} &&= S_\mathrm{abs,e} + S_\mathrm{abs,Ar+} + S_\mathrm{abs,Xe+} &&, \label{eq:tot}\\
        &S_\mathrm{abs,e} &&= 2\,( \Gamma_\mathrm{e}\,\varepsilon_\mathrm{e} + \Gamma_\mathrm{Ar+}\,\varepsilon_\mathrm{c,e,Ar} + \Gamma_\mathrm{Xe+}\,\varepsilon_\mathrm{c,e,Xe} ) &&, \label{eq:e}\\
        &S_\mathrm{abs,Ar+} &&= 2\, \Gamma_\mathrm{Ar+}\, (\varepsilon_\mathrm{i,Ar+} + \varepsilon_\mathrm{is,Ar+} + \varepsilon_\mathrm{cx,Ar+}) &&, \label{eq:Ar}\\
        &S_\mathrm{abs,Xe+} &&= 2\, \Gamma_\mathrm{Xe+}\, (\varepsilon_\mathrm{i,Xe+} + \varepsilon_\mathrm{is,Xe+} + \varepsilon_\mathrm{cx,Xe+}) &&. \label{eq:Xe}
    \end{alignat}
    For this and more complex systems, it is useful to split the total energy flux $S_\mathrm{abs,tot}$ into a separate term for each species.
    This separation is done in equation \eqref{eq:tot}.
    Besides, we split the collisional losses to the background gas for ions, $\varepsilon_\mathrm{c,i}$, into two terms.
    One represents the losses due to charge exchange collisions for Ar\textsuperscript{+} ions ($\varepsilon_\mathrm{cx,Ar+}$) and Xe\textsuperscript{+} ions ($\varepsilon_\mathrm{cx,Xe+}$).
    The other term gives the losses caused by the remaining isotropic scattering.
    It separates the isotropic losses for Ar\textsuperscript{+} ions ($\varepsilon_\mathrm{is,Ar+}$), and Xe\textsuperscript{+} ions ($\varepsilon_\mathrm{is,Xe+}$).
    This distinction is based on the nomenclature of Phelps \cite{Phelps} and will prove useful for understanding the ion dynamics.\par
    The terms for the electron flux ($\Gamma_\mathrm{e}$), the Ar\textsuperscript{+} ion flux ($\Gamma_\mathrm{Ar+}$) and the Xe\textsuperscript{+} ion flux ($\Gamma_\mathrm{Xe+}$) in this model are obtained from the PIC/MCC simulation at the surface of the electrodes.
    
%%%%%%%
\section{Results and Discussion} \label{results}    
    
%%%%%%%%%%%%%%%%%%%%%%

\subsection{Influence of the neutral gas composition on the discharge}\label{GasVar}

    \begin{figure*}[t!]
    	\begin{center}
           	\includegraphics[width = \textwidth]{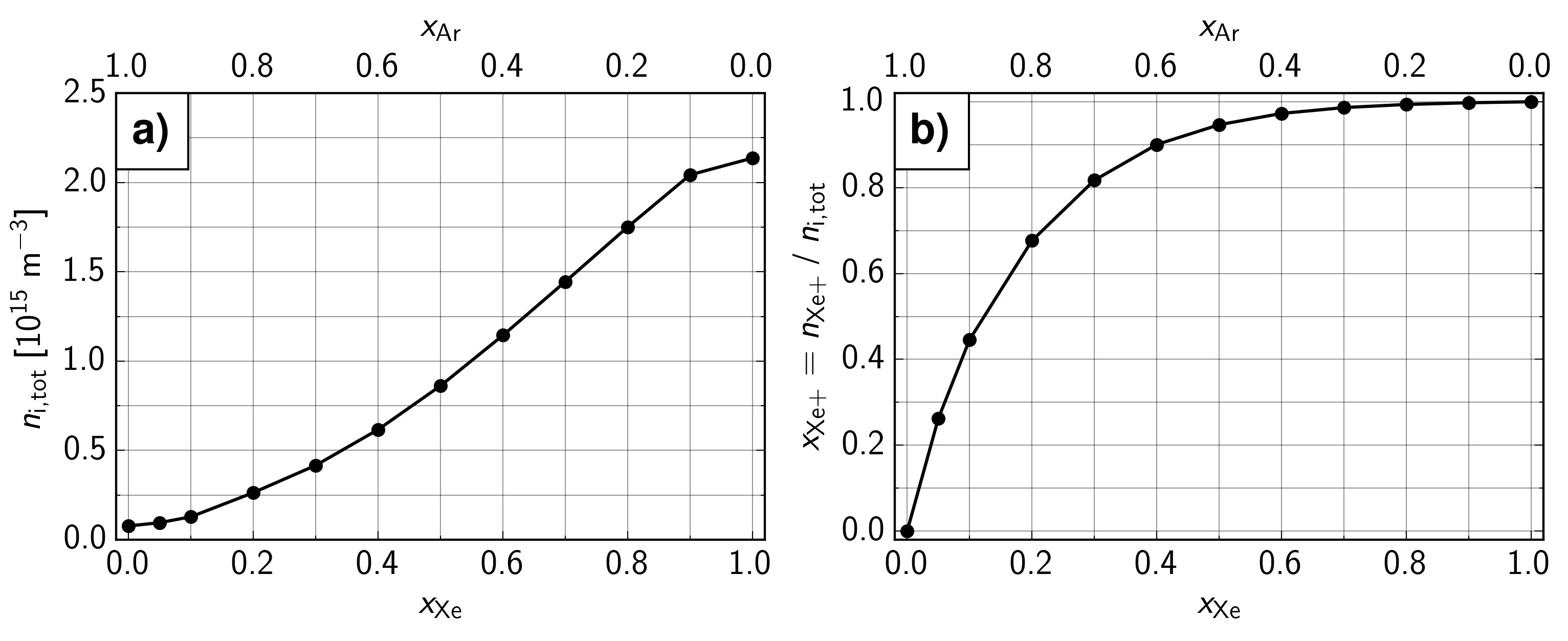}
    		\caption{The trend of the plasma density while varying the background gas composition.
			a) shows the development of the time and space averaged total ion density.
			b) shows the fraction of Xe\textsuperscript{+} ions in the discharge.
			(conditions: $p_\mathrm{gas} = 3\,$Pa, $l_\mathrm{gap} = 25\,$mm, $V_\mathrm{RF} = 100\,$V, $f_\mathrm{RF} = 13.56\,$MHz)}
    		\label{fig:ni_GasVar}
    	\end{center}
    \end{figure*}
    
    Initially, the influence of the gas composition on the discharge is investigated by varying the Ar/Xe density ratio.
    Figure \ref{fig:ni_GasVar} a) shows the total ion density $n_\mathrm{i,tot}$ as a function of the xenon gas fraction $x_\mathrm{Xe}$ or the argon gas fraction $x_\mathrm{Ar}$, respectively.
    Here, the total ion density $n_\mathrm{i,tot}$ is defined as the sum of the spatially and temporally averaged number densities of Ar\textsuperscript{+} and Xe\textsuperscript{+} ions.
    The gas fractions of argon and xenon are defined by the ratio of the respective species density and the total gas density.
    In analogy to this definition, we define an ion fraction, e.g., the fraction of Xe\textsuperscript{+} ions $x_\mathrm{Xe+}$, as the ratio of the number density of Xe\textsuperscript{+} ions and the total ion density $n_\mathrm{i,tot}$.
    Figure \ref{fig:ni_GasVar} b) depicts this Xe\textsuperscript{+} ion ratio as a function of the xenon gas fraction $x_\mathrm{Xe}$ or argon gas fraction $x_\mathrm{Ar}$, respectively.\par
    When varying the gas mixture from pure argon to pure xenon by successively increasing the xenon fraction $x_\mathrm{Xe}$, the plasma density rises significantly over about one order of magnitude (fig. \ref{fig:ni_GasVar} a)).
    The ratio of Xe\textsuperscript{+} ions (fig. \ref{fig:ni_GasVar} b)) reveals that even small admixtures of xenon to an argon gas produce a high amount of Xe\textsuperscript{+} ions.
    A xenon fraction of $x_\mathrm{Xe} \approx 0.15$ is already sufficient for Xe\textsuperscript{+} ions to become the dominant ion species.
    Xenon admixtures of about 30 percent ($x_\mathrm{Xe} = 0.3$) produce a strongly Xe\textsuperscript{+} dominated discharge ($x_\mathrm{Xe+} \gtrsim 0.8$).
    Both the development of the plasma density and the fraction of Xe\textsuperscript{+} ions in the discharge as a function of the gas composition show non-linear relations.
    Hereby, the trend of figure \ref{fig:ni_GasVar} a) approximates a compressed parabola whilst the trend of \ref{fig:ni_GasVar} b) resembles the function of the square root.
    In the following, the overall dominance of Xe\textsuperscript{+} ions will be examined and explained in more detail.
    The difference in the ionization energies gives a basic explanation of the observed behavior.
    The ionization threshold for xenon ($\varepsilon_\mathrm{thr,i,Xe} = 12.12\,$eV) is much smaller than the threshold for argon ($\varepsilon_\mathrm{thr,i,Ar} = 15.8\,$eV) (comp. tab. \ref{tab:chemistry}).
    This disparity allows lower energetic electrons to ionize xenon in contrast to argon.
    Additionally, the ionization cross section of xenon $\sigma_\mathrm{i,Xe}$ is about one order of magnitude bigger than the corresponding cross section $\sigma_\mathrm{i,Ar}$ for argon (comp. fig. \ref{fig:crosssections} a) and b)).
    As a result, Xe\textsuperscript{+} ions are prevalent, even for low xenon admixtures, and dominate the discharge for a wide mixture range.
    This result agrees with previous works \cite{mixture2, mixture4} that, for different mixtures, have shown at least one dominant ion species for a wide range of admixtures.
    
    \begin{figure*}[t!]
    	\begin{center}
            \includegraphics[width = \textwidth]{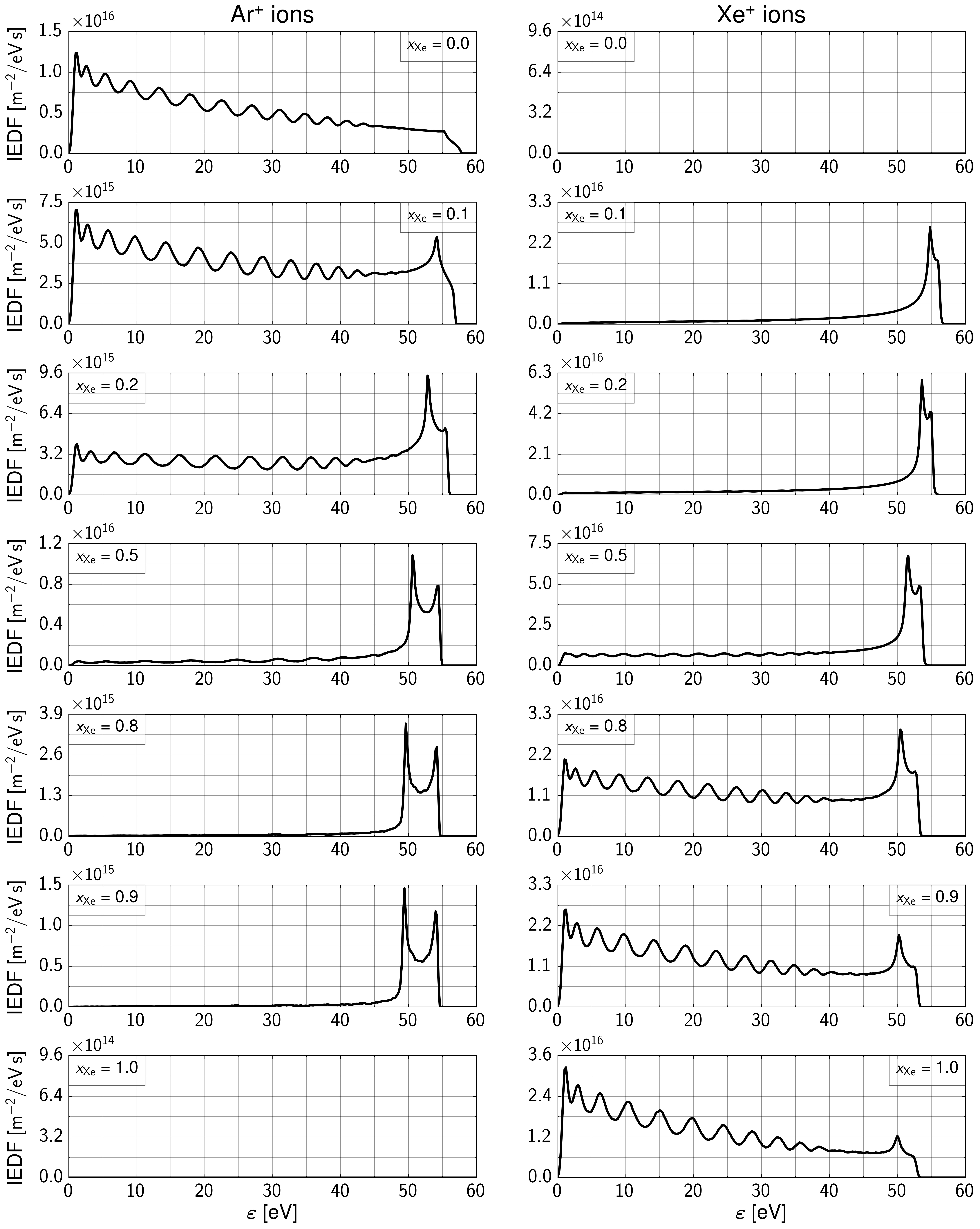}
    		\caption{Ion energy distribution function (IEDF) at the electrode for different compositions of the background gas.
			The left column shows distribution functions for Ar\textsuperscript{+} ions.
			The corresponding distributions for Xe\textsuperscript{+} ions are on the right side of the plot.
			(conditions: $p_\mathrm{gas} = 3\,$Pa, $l_\mathrm{gap} = 25\,$mm, $V_\mathrm{RF} = 100\,$V, $f_\mathrm{RF} = 13.56\,$MHz)}
    		\label{fig:IEDF_GasVar}
    	\end{center}
    \end{figure*}
    
    The influence of the gas composition also directly manifests in a variation of the IEDFs for both ion species.
    The plots of figure \ref{fig:IEDF_GasVar} show IEDFs for both Ar\textsuperscript{+} and Xe\textsuperscript{+} ions at the electrode surface.
    The energy, plotted on the abscissa, is given in eV.
    The ordinate shows the IEDF normed on the respective ion flux $\Gamma_\mathrm{i,s}$ at the electrode.
    Each row of figure \ref{fig:IEDF_GasVar} represents results for both ion species and the same case.
    The cases are distinguished by the xenon fraction $x_\mathrm{Xe}$ as indicated.
    Here, the plots in the right column show IEDFs of Ar\textsuperscript{+} ions, and the results for Xe\textsuperscript{+} ions are shown in the right column. \par
    In section \ref{chemistry}, we argue that the charge exchange between Ar\textsuperscript{+}/Xe and Xe\textsuperscript{+}/Ar, respectively, is a non-resonant process and a three-body collision.
    We conclude that this process is negligible.
    As a result, a variation of the gas composition changes the ions' probability to perform charge exchange collisions.
    Therefore, Ar\textsuperscript{+} ions, for high argon fractions $x_\mathrm{Ar}$, show an IEDF clearly dominated by collisions.
    This IEDF becomes a collisionless distribution for small argon admixtures to a xenon background (fig. \ref{fig:IEDF_GasVar} left).
    The IEDF of Xe\textsuperscript{+} ions shows a similar trend except that Xe\textsuperscript{+} ions have a less distinct bimodal behaviour for the cases with high argon fraction.
    This difference is explained by the scaling of the width of the bimodal peak being proportional to $\sqrt{m_\mathrm{i}^{-1}}$ \cite{ChabertBook, Kawamura}.
    Besides, an argon fraction $x_\mathrm{Ar}$ of 0.2 and 0.3 or, vice versa, a xenon fraction $x_\mathrm{Xe}$ of 0.2 and 0.3 creates an intermediate or hybrid regime.
    A significant number of ions experiences the discharge as being collision dominated, while the remaining ions cross the sheath collisionlessly.
    In figure \ref{fig:IEDF_GasVar}, the described regime is visible for argon at $x_\mathrm{Xe} = 0.2$ and xenon at $x_\mathrm{Xe} = 0.8$.
    Several distinct peaks are visible at low energies that stem from charge exchange collisions, and at high energies, the characteristic collisionless bimodal peak is clearly established.
    In these cases, particularly, the scaling of the bimodal peak width can be observed.
    For both cases, Xe\textsuperscript{+} ions establish a bimodal peak narrower than the bimodal peak formed by Ar\textsuperscript{+} ions.
  
%%%%%%%%%%%%%%%%%%%%%%
    
    \subsection{Revision and analysis of the energy balance model} \label{en-validation}

    \begin{figure*}[t!]
    	\begin{center}
    		\includegraphics[width = \textwidth]{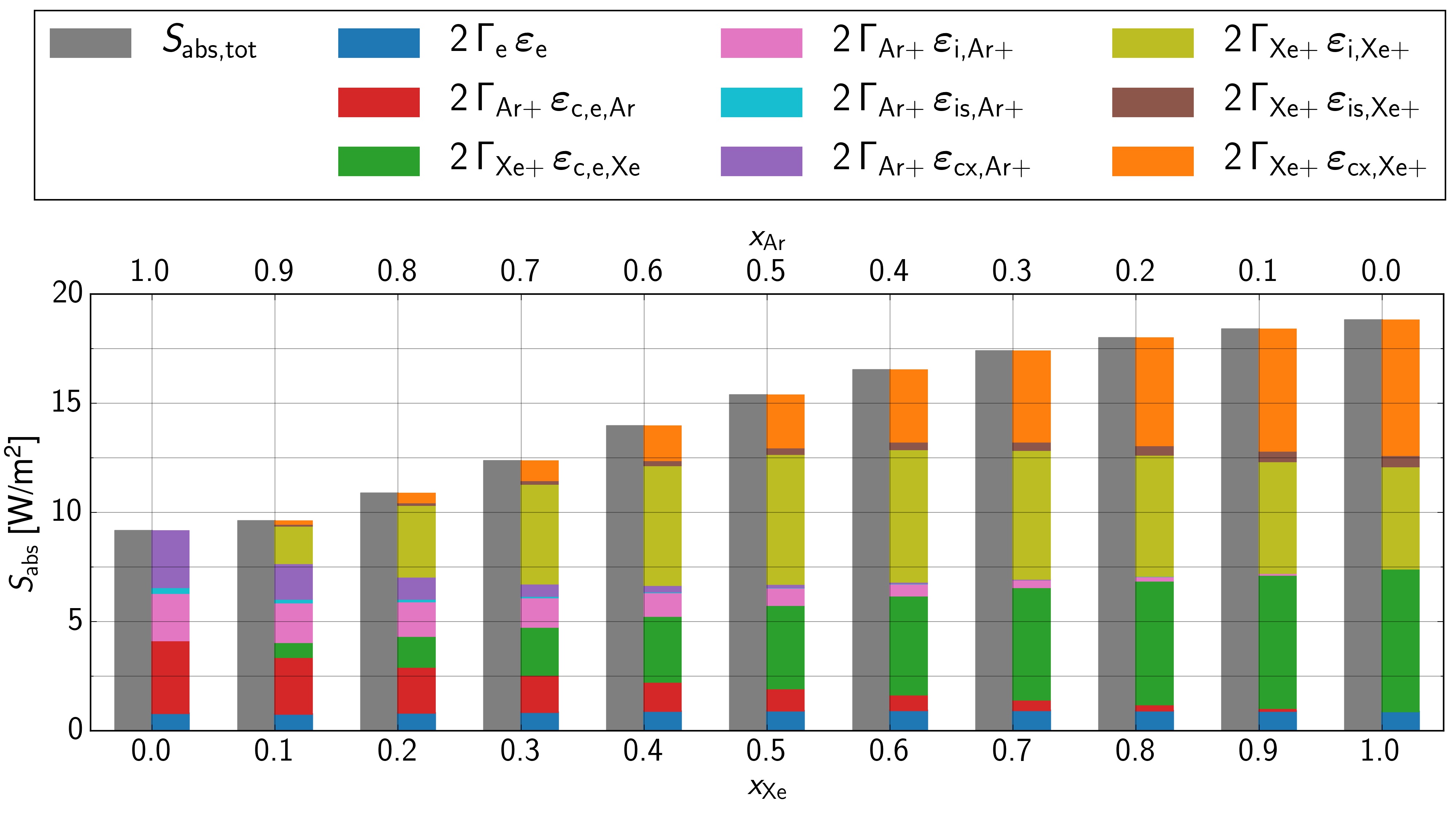}
    		\caption{Evaluation of the energy balance equations (eq. \eqref{eq:tot} - \eqref{eq:Xe}) for various gas compositions.
			All parameters have been calculated by means of a PIC/MCC simulation.
			(conditions: $p_\mathrm{gas} = 3\,$Pa, $l_\mathrm{gap} = 25\,$mm, $V_\mathrm{RF} = 100\,$V, $f_\mathrm{RF} = 13.56\,$MHz)}
    		\label{fig:confirm_eb}
    	\end{center}
    \end{figure*}
    
    For a fundamental understanding of the energy distribution within the system, the energy balance model resembled by eqs. \eqref{eq:tot} - \eqref{eq:Xe} may be used, as shown in figure \ref{fig:confirm_eb}.
    We calculate all parameters and properties by means of a PIC/MCC simulation averaged over 3000 RF periods.
    The plot shows two bars for each of the chosen gas compositions.
    The grey bar on the left-hand side represents the total absorbed energy flux $S_\mathrm{abs,tot}$.
    The colored bars on the right-hand side resolve the different channels of energy dissemination in detail.
    The colors blue (electron energy lost at the electrode $\propto\  \varepsilon_\mathrm{e}$),
    red (averaged energy consumption per e/Ar\textsuperscript{+}-pair $\propto\  \varepsilon_\mathrm{c,e,Ar}$),
    and green (averaged energy consumption per e/Xe\textsuperscript{+}-pair $\propto\  \varepsilon_\mathrm{c,e,Xe}$)  represent the right-hand side of equation \eqref{eq:e}.
    The right-hand side of equation \eqref{eq:Ar} is depicted in pink (Ar\textsuperscript{+} ion energy loss at the electrode $\propto\  \varepsilon_\mathrm{i,Ar+}$),
    cyan (energy loss by isotropic scattering $\propto\  \varepsilon_\mathrm{is,Ar+}$),
    and purple (energy loss by backscattering $\propto\  \varepsilon_\mathrm{cx,Ar+}$).
    The remaining colors olive (Xe\textsuperscript{+} ion energy loss at the electrode $\propto\  \varepsilon_\mathrm{i,Xe+}$),
    brown (energy loss by isotropic scattering $\propto\  \varepsilon_\mathrm{is,Xe+}$),
    and orange (energy loss by backscattering $\propto\  \varepsilon_\mathrm{cx,Xe+}$) visualize the right-hand-side of equation \eqref{eq:Xe}. \par
    At first, it is noticeable that figure \ref{fig:confirm_eb} shows a roughly square-root-shaped increase of the absorbed energy flux density $S_\mathrm{abs}$ as a function of the xenon fraction $x_\mathrm{Xe}$. 
    This trend is a consequence of the boundary conditions in combination with the varied gas composition.
    The PIC/MCC simulations considered in this work use a single-frequency voltage source as a boundary condition for calculating the electric field.
    The energy flux density is calculated self-consistently according to the plasma state.
    At low xenon fractions $x_\mathrm{Xe}$, xenon neutrals and Xe\textsuperscript{+} ions successively provide additional loss mechanisms, and the energy consumption increases rapidly.
    Whereas at higher xenon fractions, xenon already dominates the discharge, and the energy consumption slowly saturates.
    Lieberman and Lichtenberg present the scaling law $n_\mathrm{s} \propto\  S_\mathrm{abs}$ \cite{LiebermanBook}.
    In section \ref{GasVar}, we discussed that the trend of the plasma density $n_\mathrm{i,tot}$ as a function of the xenon fraction $x_\mathrm{Xe}$ (comp. fig. \ref{fig:ni_GasVar} a)) is approximated by a parabola.
    Combined with the square-root-shaped trend of the absorbed energy flux density $S_\mathrm{abs}$ as a function of the xenon fraction $x_\mathrm{Xe}$,
    we see the resulting trend of $n_\mathrm{i,tot}$ and $S_\mathrm{abs}$ match the anticipated scaling.\par
    The results calculated for pure argon ($x_\mathrm{Xe} = 0.0$) and pure xenon ($x_\mathrm{Xe} = 1.0$) discharges resemble the classical model given by equation \eqref{eq:cl_energybalance}.
    The results demonstrate that all individual loss terms sum up to the total energy flux and thus prove the models' exact energy conservation.
    Both the argon case and xenon case reveal that the energy loss due to colliding ions (argon: cyan and purple, xenon: brown and orange) has a significant contribution to the energy balance (argon:  $\approx 31.1\,$\% of  the total energy, xenon: $\approx 35.6\,$\%).
    These findings are similar to the study of Jiang et al.\cite{Jiang}.\par
    The remaining bars of figure \ref{fig:confirm_eb} review the modified energy balance model presented in equation \eqref{eq:tot} to \eqref{eq:Xe}.
    It shows, for some exemplary gas mixtures, that the suggested balance for multiple ion species is complete and that each species' energy transfer can be traced individually.
    Furthermore, the results show that for a complete energy balance for plasmas with two ion species, colliding ions' energy transfers are at least as important as they are in mono ionic plasmas \cite{Jiang}.
    Especially, the energy losses due to charge exchange collisions ($\varepsilon_\mathrm{cx,Ar+}$ -purple- or $\varepsilon_\mathrm{cx,Xe+}$ -orange-, resp.) make up for a significant amount of the transferred energy.\par
\begin{figure*}[t!]
    	\begin{center}
            \includegraphics[width = \textwidth]{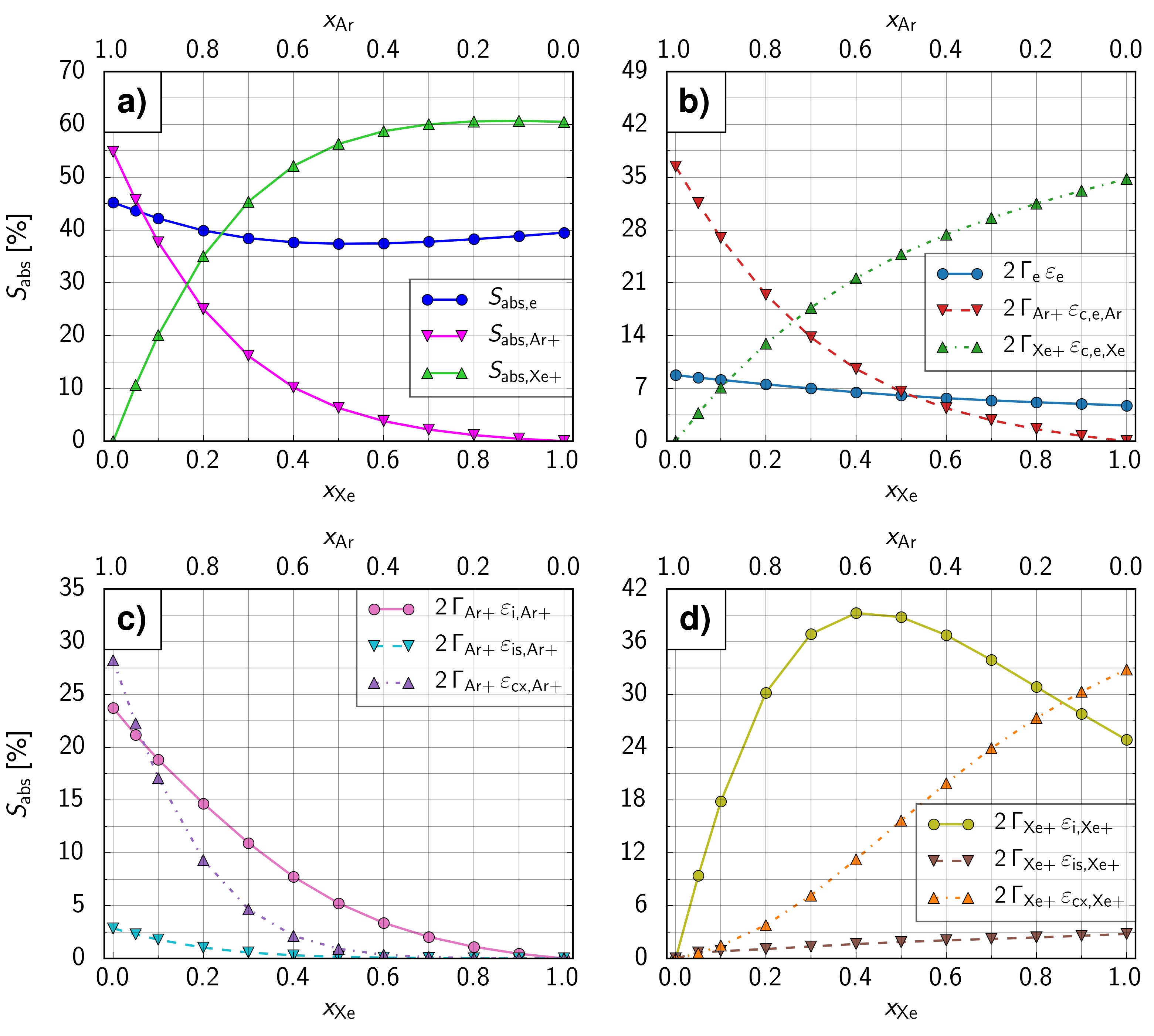}
    		\caption{The energy balance equations \eqref{eq:tot} - \eqref{eq:Xe} applied for the background gas variation.
			All properties are calculated from a PIC simulation and referred to the total absorbed energy flux $S_\mathrm{abs,tot}$.
			All plots show the right-hand side of their corresponding equation in relative units.
			a) represents equation \eqref{eq:tot}, b) equation \eqref{eq:e}, c) equation \eqref{eq:Ar} and d) equation \eqref{eq:Xe}.
			(conditions: $p_\mathrm{gas} = 3\,$Pa, $l_\mathrm{gap} = 25\,$mm, $V_\mathrm{RF} = 100\,$V, $f_\mathrm{RF} = 13.56\,$MHz)}
    		\label{fig:EB_GasVar}
    	\end{center}
    \end{figure*}

    Both the individual energy transfers of each particle species and the exact resolution of specific loss channels will in the following prove useful to understand and analyze the discharge.
    To make the results comparable, we decide to switch the representation of the energy flux density $S_\mathrm{abs}$ from absolute units to relative units (comp. fig. \ref{fig:EB_GasVar}).
    Thereby, we refer to the energy fluxes of each case individually with respect to the total energy flux $S_\mathrm{abs,tot}$.
        
    Figure \ref{fig:EB_GasVar} shows a rearrangement of the data of figure \ref{fig:confirm_eb} in the relative representation explained before.
    Each of the subplots a) to d) respectively present the right-hand side of equations \eqref{eq:tot} to \eqref{eq:Xe}.
    The abscissa of all plots mark energy flux densities in relative units, and the ordinates are in units of the gas fractions ($x_\mathrm{Xe}$ or $x_\mathrm{Ar}$, resp.).
    The color schemes for figures \ref{fig:EB_GasVar} b), c), and d) are similar to the ones used in figure \ref{fig:confirm_eb}.
    Figure \ref{fig:EB_GasVar} a) introduces a new color scheme for the total energy fluxes absorbed by electrons (bright blue), Ar\textsuperscript{+} ions (fuchsia), and Xe\textsuperscript{+} ions (lime green).\par 
    In section \ref{GasVar}, we point out two observations.
    First, Xe\textsuperscript{+} ions are for a wide range of mixtures the dominant ion species.
    Second, for constant gas pressure, collisional features of the IEDF depend on the gas composition, and even a collisional/collisionless hybrid regime can be reached.
    Both observations are confirmed and explained by the energy balance.
    Figure \ref{fig:EB_GasVar} a) shows that for a xenon fraction $x_\mathrm{Xe}$ between 0.15 and 0.2, Ar\textsuperscript{+} ions and Xe\textsuperscript{+} ions absorb an equal amount of energy ($30\, \%$ of $S_\mathrm{abs}$ or $\approx 3\,$W/m\textsuperscript{2}).
    Simultaneously, the production of Xe\textsuperscript{+} ions is more effective than the production of Ar\textsuperscript{+} ions.
    This increased effectiveness is due to the lower ionization energy of xenon ($\varepsilon_\mathrm{thr,Xe} = 12.12\, \mathrm{eV}$) compared to argon's ionization energy ($\varepsilon_\mathrm{thr,Ar} = 15.8\, \mathrm{eV}$).\par
    The case for a xenon admixture of 20 percent ($x_\mathrm{Xe} = 0.2$) serves as the best example for this finding.
    There are, with a Xe\textsuperscript{+} ion fraction $x_\mathrm{Xe+} \approx 0.7$ (comp. fig. \ref{fig:ni_GasVar}), more Xe\textsuperscript{+} ions than Ar\textsuperscript{+} ions inside the discharge.
    Nevertheless, more energy per electron-ion pair is consumed to produce Ar\textsuperscript{+} ions (red) than for the generation of Xe\textsuperscript{+} ions (green) (fig. \ref{fig:EB_GasVar} b)).
    This finding is explained by the lower excitation and ionization levels of xenon compared to argon.
    Simultaneously, these lower excitation and ionization levels open up new loss channels for the electrons inside the system.
    Raising the xenon fraction $x_\mathrm{Xe}$ yields more and more electrons that are not energetic enough to participate in inelastic processes in an argon discharge.
    Thus, the averaged electron energy $\overline{\varepsilon_\mathrm{e}}$ drops when going from an argon discharge to a xenon discharge.
    The decreasing loss term $\varepsilon_\mathrm{e}$ (blue) in figure \ref{fig:EB_GasVar} b) hints at the average electron energy of the system and gives evidence of this explanation.
    All in all, this shows that the production of Xe\textsuperscript{+} ions fills an unoccupied energetic niche where numerous low energetic electrons can participate.
    Therefore, a significant production of Xe\textsuperscript{+} ions is observed even for low xenon fractions $x_\mathrm{Xe}$ and Xe\textsuperscript{+} ions are the dominant ion species for the majority of the possible Ar/Xe mixtures. \par
    The trends observed in the IEDFs (fig. \ref{fig:IEDF_GasVar}) and the conclusions drawn from this observation are confirmed by the energy balance as well (fig. \ref{fig:EB_GasVar} c) and d)).
    Looking at the losses due to charge exchange collisions $\varepsilon_\mathrm{cx,i}$ for both Ar\textsuperscript{+} ions (fig. \ref{fig:EB_GasVar} c), purple) and Xe\textsuperscript{+} ions (fig. \ref{fig:EB_GasVar} d), orange), it becomes apparent that the collisional features are switched between Ar\textsuperscript{+} and Xe\textsuperscript{+} ions when going towards more argon, or xenon respectively, dominated gas mixtures.
    The losses due to charge exchange for Ar\textsuperscript{+} ions $\varepsilon_\mathrm{cx,Ar+}$ monotonically fall as a function of the xenon fraction $x_\mathrm{Xe}$ (fig. \ref{fig:EB_GasVar} c), purple) while the corresponding term for Xe\textsuperscript{+} ions $\varepsilon_\mathrm{cx,Xe+}$ monotonically raises, when displayed as the same relation (fig. \ref{fig:EB_GasVar} d), orange).
    The slight difference in the trends is explained by the dominance of the Xe\textsuperscript{+} ions in the discharge.
    While the density of Xe\textsuperscript{+} ions rapidly increases, when adding small amounts of xenon to an argon background (comp. fig. \ref{fig:ni_GasVar} a)), the density of Ar\textsuperscript{+} ions vanishes as fast among the dominant Xe\textsuperscript{+} ions.
    Hence, there are not enough Ar\textsuperscript{+} ions present in discharges dominated by Xe\textsuperscript{+} ions, so that the losses of Ar\textsuperscript{+} ions in total cannot significantly contribute to the energy absorbed by the discharge (fig. \ref{fig:EB_GasVar} a), fuchsia).\par
    In addition to this, the mean energy of Xe\textsuperscript{+} ions at the electrode $\varepsilon_\mathrm{i,Xe+}$ shows a very different trend than all the collisional quantities (fig. in \ref{fig:EB_GasVar} d), olive).
    Instead of monotonically rising with the xenon fraction $x_\mathrm{Xe}$ as the corresponding Ar\textsuperscript{+} term does as a function of the argon ratio $x_\mathrm{Ar}$ (comp. fig. \ref{fig:EB_GasVar} c), pink), the Xe\textsuperscript{+} curve shows a maximum at $x_\mathrm{Xe} = 0.4$.
    This maximum is closely connected to the dominance of Xe\textsuperscript{+} ions.
    At 40 percent xenon admixture ($x_\mathrm{Xe} = 0.4$), Xe\textsuperscript{+} ions already make up for about 90 percent of the ions in the discharge (fig. \ref{fig:ni_GasVar} a)).
    At the same time, argon is the dominant background rendering Xe\textsuperscript{+} ions more or less incapable of doing a relevant amount of charge exchange collisions.
    This lack of charge exchange collisions is seen in the IEDF of Xe\textsuperscript{+} ions, that even for a xenon fraction $x_\mathrm{Xe} = 0.5$ shows a characteristic collisionless single bimodal peak (fig. \ref{fig:IEDF_GasVar}, right).
    For lower xenon fractions $x_\mathrm{Xe}$, the number density $n_\mathrm{Xe+}$ and the flux density $\Gamma_\mathrm{Xe+}$ are lower and fewer Xe\textsuperscript{+} ions reach the electrode.
    This decrease results in a lower energy loss.
     For higher xenon fractions $x_\mathrm{Xe}$, the charge transfer collision of Xe/Xe\textsuperscript{+} becomes more and more probable.
     This trend manifests in the IEDFs (fig. \ref{fig:IEDF_GasVar}, right) and the trend of the loss term for charge exchange $\varepsilon_\mathrm{cx,Xe+}$ (fig. \ref{fig:EB_GasVar} d), orange).
     Thus, the energy loss of Xe\textsuperscript{+} ions to the surface finally drops because the energy gets dissipated more strongly to the neutral gas via charge exchange collisions. \par
    The minimum of the total energy flux density absorbed by electrons $S_\mathrm{abs,e}$ (fig. \ref{fig:EB_GasVar} a), bright blue) has a similar explanation.
    For a xenon fraction $x_\mathrm{Xe} = 0.5$, electrons absorb the lowest amount of energy.
    Under these conditions, Xe\textsuperscript{+} ions make up for almost all the ions in the discharge.
    Figure \ref{fig:ni_GasVar} b) shows that for a xenon fraction $x_\mathrm{Xe} = 0.5$ the Xe\textsuperscript{+} ion fraction $x_\mathrm{Xe+}$ is approximately 0.9.
    At the same time, xenon atoms make up for just 50 percent of the background gas.
    The amount of collisions with argon or xenon particles respectively is, as argued before, significantly reduced compared to mixtures with a high amount of either of the gases.
    Thus, for xenon fractions $x_\mathrm{Xe} < 0.5$, the production of Ar\textsuperscript{+} ions causes electrons to absorb and invest more energy.
    For xenon fraction $x_\mathrm{Xe} > 0.5$, collisions with xenon neutrals become successively more probable, and the production of Xe\textsuperscript{+} ions consumes more energy (comp. fig. \ref{fig:EB_GasVar} b), green) without significantly changing the discharge conditions any more (comp. fig. \ref{fig:ni_GasVar}).\par
    Additionally, figure \ref{fig:IEDF_GasVar} shows that $x_\mathrm{Xe} = 0.5$ is optimal for producing high energetic ions.
    Both ion species establish the characteristic collisionless bimodal peaks and impact the surface with high energies.
    Therefore, the relative amount of energy brought by ions to the surface is maximal.
    For lower xenon fraction ($x_\mathrm{Xe} < 0.5$), the IEDF of Ar\textsuperscript{+} ions is visibly affected by collisions and vice versa for higher xenon fraction ($x_\mathrm{Xe} > 0.5$).
    
%%%%%%%%%%%%%%%%%%%%%%%%%
    
\subsection{Influence of the driving voltage on the discharge} \label{VolVar}

    \begin{figure*}[t!]
    	\begin{center}
            \includegraphics[width = \textwidth]{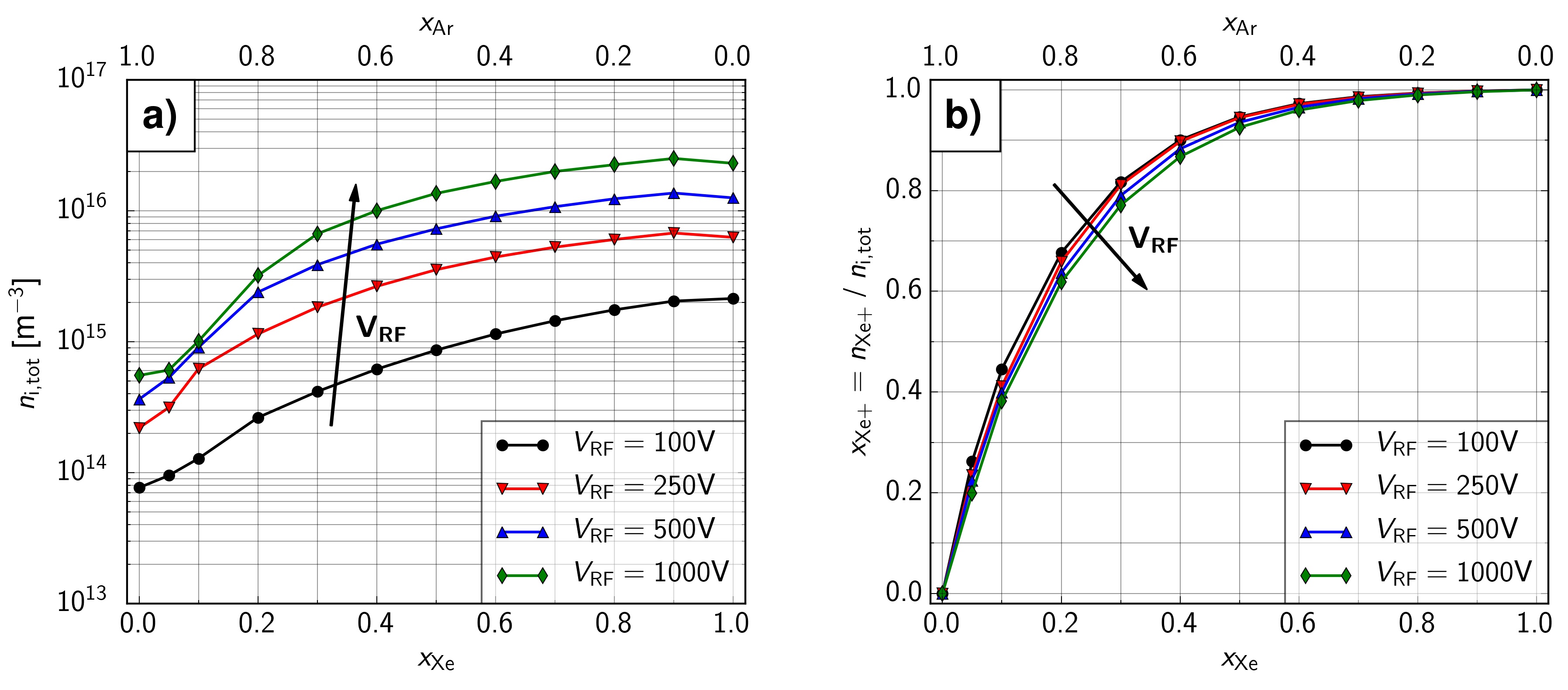}
    		\caption{The trend of the plasma density while varying the background gas composition and driving voltage.
			a) shows the development of the time and space averaged total ion density.
			b) shows the relative fraction of Xe\textsuperscript{+} ions in the discharge.
			(conditions: $p_\mathrm{gas} = 3\,$Pa, $l_\mathrm{gap} = 25\,$mm, $f_\mathrm{RF} = 13.56\,$MHz)}
    		\label{fig:ni_VolVar}
    	\end{center}
    \end{figure*}
    
     In terms of our simulation, a raised driving voltage equals, if all other parameters (gas composition, pressure, etc.) are kept constant, raising energy input to the system.
     Figure \ref{fig:ni_VolVar} a) shows a semi-logarithmic representation of the time and space averaged total plasma density $n_\mathrm{i,tot}$ as a function of the gas fractions ($x_\mathrm{Xe}$ or $x_\mathrm{Ar}$, resp.).
     The different colors differentiate the data for different RF amplitudes (black = $100\,$V, red = $250\,$V, blue = $500\,$V, green = $1\,$kV).
     The black curve shows the same data as figure \ref{fig:ni_GasVar} a).
     Due to the aforementioned higher input energy, the plasma density is raised in general while the several curves' general trend is kept.
     Independent of the driving voltage, argon discharges have a significantly lower plasma density than xenon discharges, and the transition while varying the gas composition shows the same non-linear trend.
     In sections \ref{GasVar} and \ref{en-validation}, we discuss that in this context, non-linear means parabolic. \par
    Apart from this, a varied driving voltage alters the dominance of Xe\textsuperscript{+} ions.
    Figure \ref{fig:ni_VolVar} b) shows a similar plot to figure \ref{fig:ni_GasVar} b).
    The Xe\textsuperscript{+} ion fraction is presented as a function of the gas fraction ($x_\mathrm{Xe}$ or $x_\mathrm{Ar}$, resp.).
    The colors have the same meanings as in figure \ref{fig:ni_VolVar} a), and the black curve was also presented before (see fig. \ref{fig:ni_GasVar} b)).
    Figure \ref{fig:ni_VolVar} b) shows that, for a fixed xenon fraction $x_\mathrm{Xe}$, a raised voltage reduces the fraction of Xe\textsuperscript{+} ions $x_\mathrm{Xe+}$ present in the discharge.
    The case for $x_\mathrm{Xe} = 0.2$ is a good example of this observation.
    When increasing the driving voltage from 100$\,$V to 1$\,$kV, the ratio of Xe\textsuperscript{+} ions $x_\mathrm{Xe+}$ drops from approximately 0.7 to roughly 0.6.

    \begin{figure*}[t!]
    	\begin{center}
            \includegraphics[width = \textwidth]{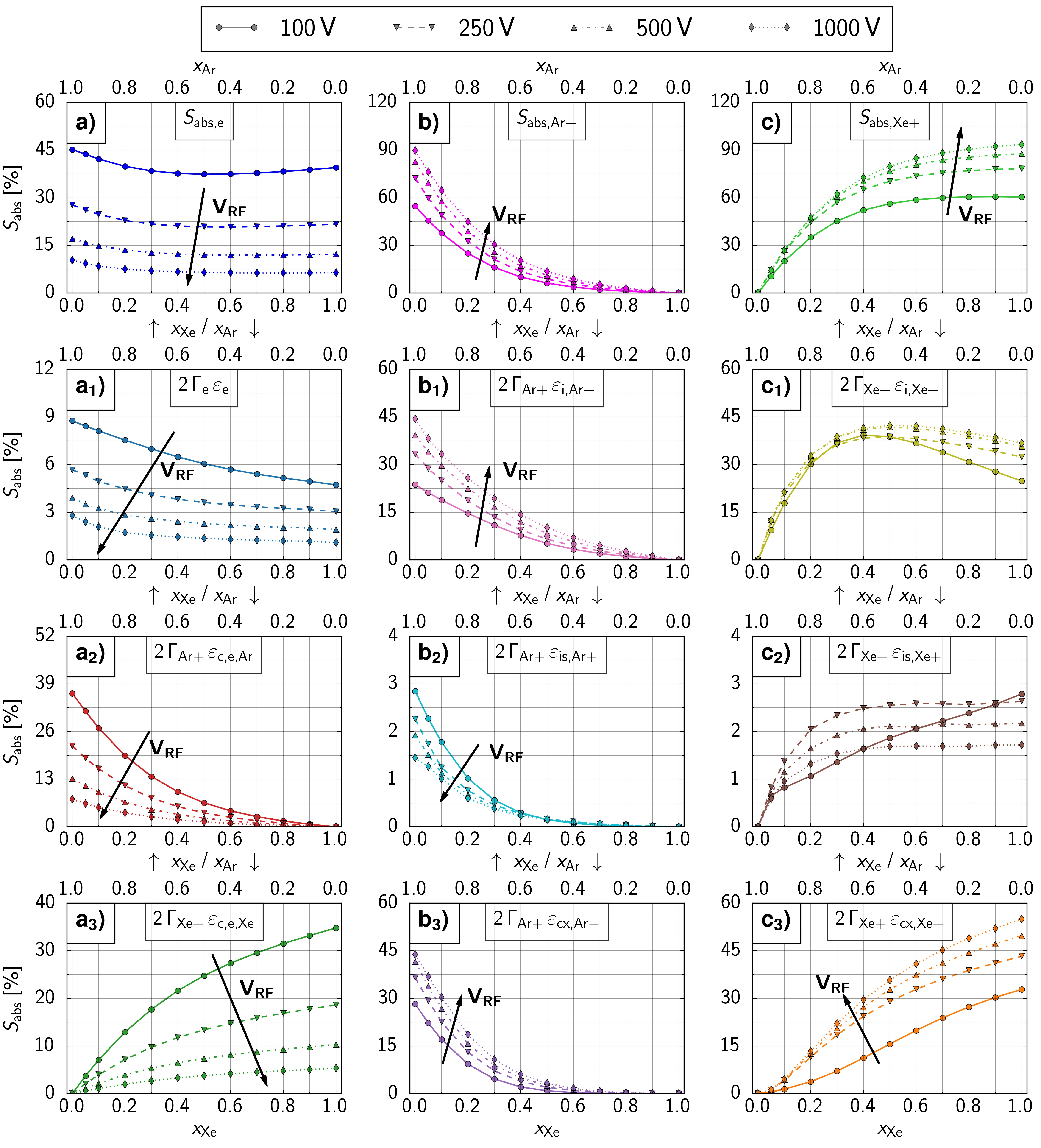}
    		\caption{The energy balance equations \eqref{eq:tot} - \eqref{eq:Xe} applied for both the variation of the background gas and the driving voltage $V_\mathrm{RF}$.
			All properties are calculated from a PIC simulation and referred to the total absorbed energy flux $S_\mathrm{abs,tot}$.
			All plots show one term on the right-hand side of their corresponding equation in relative units.
			a) shows the electron's part $S_\mathrm{abs,e}$ of eq. \eqref{eq:tot}. a\textsubscript{1}) - a\textsubscript{3}) show the three terms of equation \eqref{eq:e} and add up to the respective curve of a).
			b) represents the Ar\textsuperscript{+} ions' part $S_\mathrm{abs,Ar+}$ of equation \eqref{eq:tot}. b\textsubscript{1}) - b\textsubscript{3}) present the three terms of equation \eqref{eq:Ar} and sum up to the respective curve of b).
			c) shows the Xe\textsuperscript{+} ions' part $S_\mathrm{abs,Xe+}$ of equation \eqref{eq:tot}. c\textsubscript{1}) - c\textsubscript{3}) depict the three terms of equation \eqref{eq:Xe}, and their addition gives the respective curve of c).
			(conditions: $p_\mathrm{gas} = 3\,$Pa, $l_\mathrm{gap} = 25\,$mm, $f_\mathrm{RF} = 13.56\,$MHz)}
    		\label{fig:EBQs_VolVar}
    	\end{center}
    \end{figure*}
    
    Once again, the energy balance (fig. \ref{fig:EBQs_VolVar}) explains the discharge mechanisms governing how an increased driving voltage raises the plasma density.
    Similar to figure \ref{fig:EB_GasVar}, terms on the right-hand side of the energy balance equations \eqref{eq:tot} - \eqref{eq:Xe} are shown in relative units and as a function of the gas fractions ($x_\mathrm{Xe}$ or $x_\mathrm{Ar}$, resp.).
    In contrast to figure \ref{fig:EB_GasVar}, each panel of figure \ref{fig:EBQs_VolVar} represents just one term of the respective equation's right-hand side.
    The different curves represent data for different driving voltages $V_\mathrm{RF}$, ranging from $V_\mathrm{RF} = 100\,$V to $V_\mathrm{RF} = 1000\,$V.
    The color scheme is analog to figures \ref{fig:confirm_eb} and \ref{fig:EB_GasVar}.
    Figure \ref{fig:EBQs_VolVar} a) shows the total energy flux density absorbed by electrons ($S_\mathrm{abs,e}$) in bright blue.
    Figure \ref{fig:EBQs_VolVar} b) depicts the total energy flux density absorbed by Ar\textsuperscript{+} ions ($S_\mathrm{abs,Ar+}$) in fuchsia, and figure \ref{fig:EBQs_VolVar} c) presents the total energy flux density absorbed by Xe\textsuperscript{+} ions ($S_\mathrm{abs,Xe+}$) in lime green.
    Together figures \ref{fig:EBQs_VolVar} a) - c) show the right-hand side of equation \eqref{eq:tot}.
    Therefore, the corresponding data points horizontally always add up to give $100\,\%$ (or the total energy flux density $S_\mathrm{abs,tot}$, resp.).
    Vertically the details of each particle species' power absorption are presented.
    Figures \ref{fig:EBQs_VolVar} a\textsubscript{1}) - a\textsubscript{3}) each show one term of the right-hand side of equation \eqref{eq:e}.
    The average energy loss of electrons at the electrodes $\varepsilon_\mathrm{e}$ is shown in figure \ref{fig:EBQs_VolVar} a\textsubscript{1}) in blue.
    The averaged amount of energy needed to create an electron/Ar\textsuperscript{+} ion pair ($\varepsilon_\mathrm{c,e,Ar}$) is found in panel a\textsubscript{2}) in red, and the related term for electron/Xe\textsuperscript{+} ion pairs ($\varepsilon_\mathrm{c,e,Xe}$) is depicted in panel a\textsubscript{3}) in green.
    The individual terms of the right-hand side of equation \eqref{eq:Ar} are shown in figures \ref{fig:EBQs_VolVar} b\textsubscript{1}) - b\textsubscript{3}).
    They reveal the details of the Ar\textsuperscript{+} ion dynamics by presenting the average energy loss by Ar\textsuperscript{+} ions at the electrodes ($\varepsilon_\mathrm{i,Ar+}$, fig. \ref{fig:EBQs_VolVar} b\textsubscript{1}), pink), the energy loss of Ar\textsuperscript{+} ions caused by isotropic scattering ($\varepsilon_\mathrm{is,Ar+}$, fig. \ref{fig:EBQs_VolVar} b\textsubscript{2}), cyan), and the energy loss of Ar\textsuperscript{+} ions due to backscattering ($\varepsilon_\mathrm{cx,Ar+}$, fig. \ref{fig:EBQs_VolVar} b\textsubscript{3}), purple).
    Similarly, figures \ref{fig:EBQs_VolVar} c\textsubscript{1}) - c\textsubscript{3}) show the right-hand side of equation \eqref{eq:Xe}.
    They unravel the details of the Xe\textsuperscript{+} ion dynamics by showing the average impingement energy of Xe\textsuperscript{+} ions at the electrodes ($\varepsilon_\mathrm{i,Xe+}$, fig. \ref{fig:EBQs_VolVar} c\textsubscript{1}), olive), the energy lost by Xe\textsuperscript{+} ions in isotropic scattering collisions ($\varepsilon_{is,Xe+}$, fig. \ref{fig:EBQs_VolVar} c\textsubscript{2}), brown), and the energy lost by Xe\textsuperscript{+} ion in backscattering collisions ($\varepsilon_{cx,Xe+}$, fig. \ref{fig:EBQs_VolVar} c\textsubscript{3}), orange).
    Vertically, the sum of the data in the subscript labeled panels gives the curves of the non-subscript labeled one (e.g., panels a\textsubscript{1}) - a\textsubscript{3}) sum-up to panel a)).\par
    In general, it is apparent that a raised driving voltage reduces the ratio of energy coupled to the electrons (fig. \ref{fig:EBQs_VolVar} a)) and raises the fraction absorbed by both Ar\textsuperscript{+} and Xe\textsuperscript{+} ions (fig. \ref{fig:EBQs_VolVar} b) or fig. \ref{fig:EBQs_VolVar} c), resp.).
    The increased energy consumption into the ion contribution mainly consists of two parts.
    First, a raised driving voltage $V_\mathrm{RF}$ increases the voltage drop across the boundary sheaths, and ions gain higher impingement energies after crossing the sheath collisionlessly.
    This is shown in figure \ref{fig:EBQs_VolVar} b\textsubscript{1}) for Ar\textsuperscript{+} ions and in figure \ref{fig:EBQs_VolVar} c\textsubscript{1}) for Xe\textsuperscript{+} ions.
    Second, an increased energy gain for the ions inside the sheath goes along with an increased energy loss caused by charge exchange collisions.
    The corresponding terms $\varepsilon_\mathrm{cx,Ar+}$ for Ar\textsuperscript{+} ions (fig. \ref{fig:EBQs_VolVar} b\textsubscript{3})) and $\varepsilon_\mathrm{cx,Xe+}$ for Xe\textsuperscript{+} ions (fig. \ref{fig:EBQs_VolVar} c\textsubscript{3})) support this hypothesis.
    Furthermore, the cross sections for charge exchange dominate the ones for isotropic scattering at high energies (comp. fig. \ref{fig:crosssections} c) and d)).
    Correspondingly, the already low energy losses by Ar\textsuperscript{+} ions ($\varepsilon_\mathrm{is,Ar+}$, fig. \ref{fig:EBQs_VolVar} b\textsubscript{2})) and Xe\textsuperscript{+} ions ($\varepsilon_\mathrm{is,Xe+}$, fig. \ref{fig:EBQs_VolVar} c\textsubscript{2})) caused by isotropic scattering decrease due to the increased driving voltage $V_\mathrm{RF}$.
    The maximum of figure \ref{fig:EBQs_VolVar} c\textsubscript{1}) was discussed in section \ref{en-validation}.
    The energy-efficient production of Xe\textsuperscript{+} ions already creates a high amount of Xe\textsuperscript{+} ions for small xenon fractions $x_\mathrm{Xe}$.
    Thus, there are optimal parameters for Xe\textsuperscript{+} ions to bombard the surface with the least collisional loss ($x_\mathrm{Xe} = 0.4$ for $V_\mathrm{RF} = 100\,$V, sec. \ref{en-validation}).
    The aforementioned enhanced role of backscattering and decreased influence of isotropic scattering causes the optimal parameters for higher driving voltages $V_\mathrm{RF}$ to shift to higher xenon fractions $x_\mathrm{Xe}$ (e.g., $X_\mathrm{Xe} = 0.5$ for $V_\mathrm{RF} = 1000\,$V, fig. \ref{fig:EBQs_VolVar} c\textsubscript{1})). \par
     In terms of ion production, the previous assessment shows that the higher the driving voltage is set, the smaller the fraction of the energy consumed for creating new electron/ion pairs becomes.
     Furthermore, the maximal amount of energy consumed for creating Xe\textsuperscript{+} ions in a pure xenon background ($x_\mathrm{Xe} = 1.0$, fig. \ref{fig:EBQs_VolVar} a\textsubscript{3})) is always lower than the corresponding maximum for Ar\textsuperscript{+} ions in a pure argon background ($x_\mathrm{Ar} = 1.0$, fig. \ref{fig:EBQs_VolVar} a\textsubscript{2})).
     As argued before, this finding correlates with the fact that the threshold energies of all inelastic processes involving xenon are significantly lower than those involving argon.
     This observation additionally reveals why Xe\textsuperscript{+} ions dominate the discharge for most conditions.
     It becomes best visible by comparing the pure argon case ($x_\mathrm{Xe}=0.0$, fig. \ref{fig:EBQs_VolVar} a\textsubscript{2})) with the pure xenon case ($x_\mathrm{Xe}=1.0$, fig. \ref{fig:EBQs_VolVar} a\textsubscript{3})) for a driving voltage of 1$\,$kV.
     Here, roughly eight percent of the total energy flux density is used to produce an electron/Ar\textsuperscript{+} ion pair (fig. \ref{fig:EBQs_VolVar} a\textsubscript{2})).
     For the corresponding pure xenon case, only five percent of the total energy is used to produce an electron/Xe\textsuperscript{+} ion pair (fig. \ref{fig:EBQs_VolVar} a\textsubscript{3})).
     Simultaneously, the xenon case's plasma density is more than one order of magnitude higher than in the argon case (comp. fig. \ref{fig:ni_VolVar} a)).\par
    
    \begin{figure*}[t!]
    	\begin{center}
            \includegraphics[width = \textwidth]{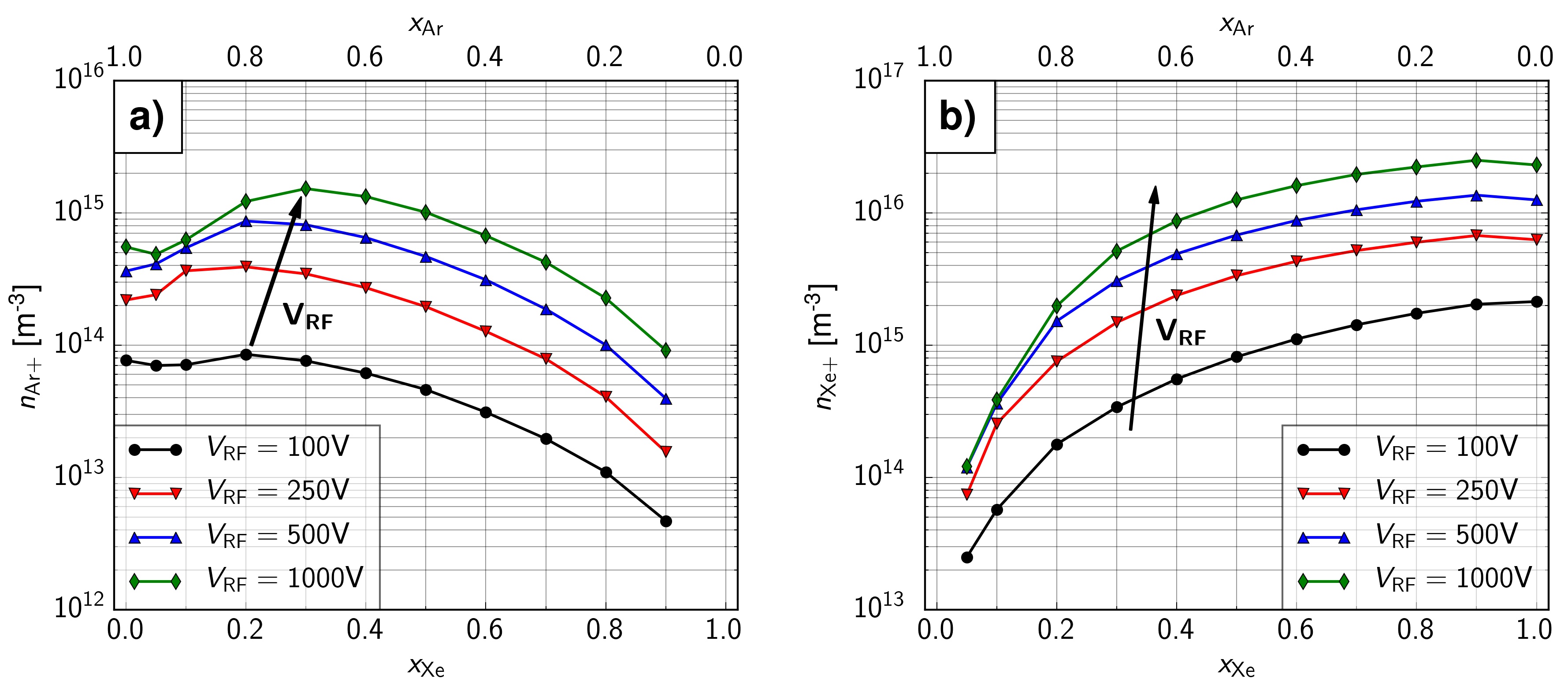}
    		\caption{The trend of the ion densities while varying the background gas composition and driving voltage.
			a) shows the development of the time and space averaged Ar\textsuperscript{+} ion density.
			b)  shows the development of the time and space averaged Xe\textsuperscript{+} ion density.
			(conditions: $p_\mathrm{gas} = 3\,$Pa, $l_\mathrm{gap} = 25\,$mm, $f_\mathrm{RF} = 13.56\,$MHz)}
    		\label{fig:ind_den_VolVar}
    	\end{center}
    \end{figure*}
    
    Since the production of Xe\textsuperscript{+} ions remains more effective for all applied driving voltages, there has to be another reason why the dominance of Xe\textsuperscript{+} ions is reduced.
    A close examination of figures \ref{fig:EBQs_VolVar} a\textsubscript{1}) and \ref{fig:ind_den_VolVar} explains the observed.
    Both panels of figure \ref{fig:ind_den_VolVar} are similar in structure to figure \ref{fig:ni_VolVar} a), but show the individual ion densities ($n_\mathrm{Ar+}$ in fig. \ref{fig:ind_den_VolVar} a) or $n_\mathrm{Xe+}$ in fig. \ref{fig:ind_den_VolVar} b), resp.) as a function of the gas fraction ($x_\mathrm{Xe}$ or $x_\mathrm{Ar}$, resp.).
    The different colors again mark different values of the driving voltage $V_\mathrm{RF}$, and the color scheme is the same as in figure \ref{fig:ni_VolVar} a).
    The trend of the Ar\textsuperscript{+} ion density in figure \ref{fig:ind_den_VolVar} a) already reveals the underlying process responsible for the decreased dominance of Xe\textsuperscript{+} ions.
    Even for the base case ($V_\mathrm{RF} =  100\,$V), the maximum of the density of Ar\textsuperscript{+} ions is found for a xenon fraction $x_\mathrm{Xe}=0.2$ and not for a xenon fraction $x_\mathrm{Xe} = 0.0$ as it is vice versa the case for Xe\textsuperscript{+} ions (comp. fig. \ref{fig:ind_den_VolVar}).
    This maximum is shifted by a raised driving voltage to a xenon admixture of 30 percent ($x_\mathrm{Xe} = 0.3$, fig. \ref{fig:ind_den_VolVar} a)).
    Recalling figure \ref{fig:ni_VolVar} a), it was observed that adding xenon to an argon background is equivalent to monotonically raising the plasma density.
    Therefore, a small xenon admixture to an argon discharge means creating more electrons that will mostly collide with an argon atom.
    As a result, the probability of ionization of an argon atom is higher than it is in a case with no or lower xenon admixture, that is, without these additional electrons.
    Thus, the density of Ar\textsuperscript{+} ions is higher than in a discharge without xenon admixture.
    This effect affects the Ar\textsuperscript{+} ions for low voltages as long as most neutrals are argon atoms.
    A raised driving voltage shifts the maximum of the Ar\textsuperscript{+} ion density and the benefits of this synergy effect to higher xenon fractions.\par
    For Xe\textsuperscript{+} ions, on the other hand, this synergy effect cannot be observed (fig. \ref{fig:ind_den_VolVar} b)).
    This observation is due to the higher ionization energy of argon.
    Figure \ref{fig:EBQs_VolVar} a\textsubscript{1}) helps to understand this observation by showing the energy lost by electrons at the electrodes $\varepsilon_\mathrm{e}$.
    The general trend of the curves for $\varepsilon_\mathrm{e}$ is a reduction by a raised driving voltage (fig. \ref{fig:EBQs_VolVar} a\textsubscript{1})).
    An equivalent conclusion is that energy is dissipated more efficiently inside the volume of the discharge.
    In terms of our non-equilibrium low-pressure discharge, there are just two ways for electrons to lose energy.
    Either they interact inelastically with the background or transfer their energy to the surface by arriving at the electrodes.
    The first option was discussed before (fig. \ref{fig:EBQs_VolVar} a\textsubscript{2}) and a\textsubscript{3})), and the second option is discussed here.
    Both processes similarly respond to the increased driving voltage, which means that a higher driving voltage increases the ion production efficiency.
    Simultaneously, the energy dissemination efficiency is increased the more xenon is added to the background gas.
    In section \ref{GasVar}, we discuss that an increased amount of xenon atoms in the discharge provides lower energetic electrons with the opportunity to get involved in inelastic processes compared to a discharge with lower or no xenon addition (see fig. \ref{fig:EB_GasVar} d)).
    In figure \ref{fig:EBQs_VolVar} a\textsubscript{1}), the same trend is observed for all depicted driving voltages.
    As a function of the xenon fraction $x_\mathrm{Xe}$, the energy lost by electrons at the electrode $\varepsilon_\mathrm{e}$ is monotonically falling.
    Vice versa, argon has higher thresholds for inelastic processes, especially ionization, than xenon (comp. tab. \ref{tab:chemistry}).
    Thus, adding argon to a xenon background cannot produce a higher electron density that would cause more ionization of xenon.
    The synergy effect does not take place for Xe\textsuperscript{+} ions that benefit from additional ionization of argon.
    
    \begin{figure*}[t!]
    	\begin{center}
            \includegraphics[width = \textwidth]{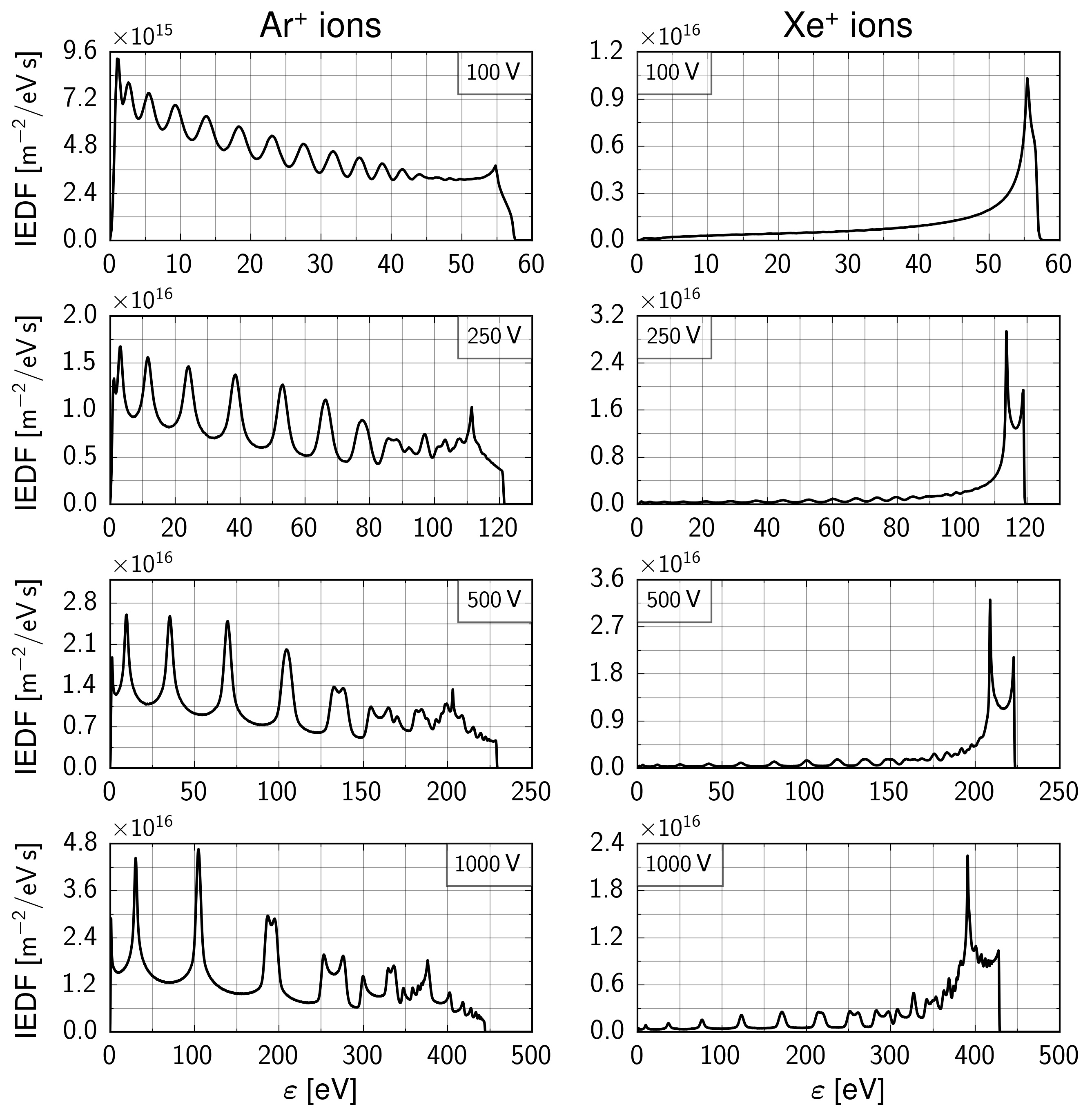}
    		\caption{Ion energy distribution function (IEDF) at the electrode for different driving voltages.
			The left column shows distribution functions for Ar\textsuperscript{+} ions.
			The corresponding distributions for Xe\textsuperscript{+} ions are on the right side of the plot.
			(conditions: $p_\mathrm{gas} = 3\,$Pa, $x_\mathrm{Xe} = 0.1$, $l_\mathrm{gap} = 25\,$mm, $f_\mathrm{RF} = 13.56\,$MHz)}
    		\label{fig:IEDF_VolVar10}
    	\end{center}
    \end{figure*}
    
    \begin{figure*}[t!]
    	\begin{center}
            \includegraphics[width = \textwidth]{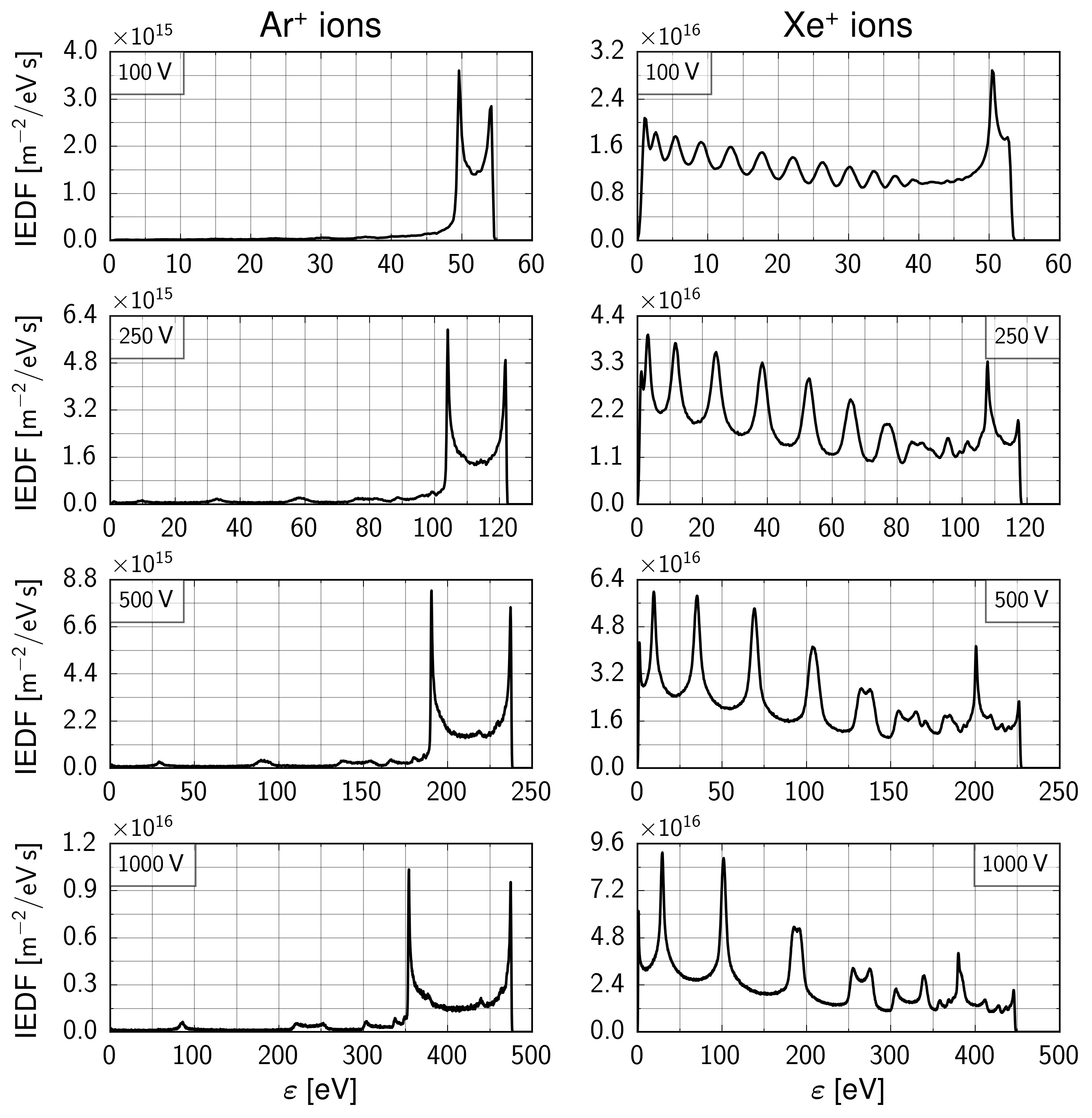}
    		\caption{Ion energy distribution function (IEDF) at the electrode for different driving voltages.
			The left column shows distribution functions for Ar\textsuperscript{+} ions.
			The corresponding distributions for Xe\textsuperscript{+} ions are on the right side of the plot.
			(conditions: $p_\mathrm{gas} = 3\,$Pa, $x_\mathrm{Xe} = 0.9$, $l_\mathrm{gap} = 25\,$mm, $f_\mathrm{RF} = 13.56\,$MHz)}
    		\label{fig:IEDF_VolVar90}
    	\end{center}
    \end{figure*}
    
    Figures \ref{fig:IEDF_VolVar10} and \ref{fig:IEDF_VolVar90} show, similar to figure \ref{fig:IEDF_GasVar}, IEDFs normalized to the respective particle flux densities at the electrode surface.
    The difference between figure \ref{fig:IEDF_VolVar10} and \ref{fig:IEDF_VolVar90} is in the gas composition (fig. \ref{fig:IEDF_VolVar10}: $x_\mathrm{Xe} = 0.1$, fig. \ref{fig:IEDF_VolVar90} $x_\mathrm{Xe} = 0.9$).
    Both figures contain IEDFs for Ar\textsuperscript{+} ions in the left column panels and IEDFs for Xe\textsuperscript{+} ions in the right one.
    The difference between each figure's four rows is the altered amplitude of the RF voltage $V_\mathrm{RF}$ given on each panel's top.
    Panels of the same row share the same voltage. \par
    The figures show that for the IEDF, a raised driving voltage, first of all, means that the averaged sheath voltage $\langle \phi_\mathrm{s} \rangle$ increases.
    This increase manifests in the width of the characteristic collisionless single bimodal peak.
    Its width scales with V$_\mathrm{RF}$ and $\sqrt{\langle \phi_\mathrm{s}(t) \rangle }\,  \mathbf{/}\,  \langle s(t) \rangle$ \cite{Kawamura, Benoit-Cattin}.
    Kawamura et al. \cite{Kawamura} give the averaged sheath width $\langle \mathrm{s}(t) \rangle$ in terms of the collisionless Child-Langmuir law:
    \begin{align}
    	\langle \mathrm{s}(t) \rangle = \frac{2}{3}\, \left( \frac{2\, e}{m_\mathrm{i}} \right)^{1/2} \, \left( \frac{\varepsilon_0}{\langle j_\mathrm{i}(t) \rangle} \right)\, \langle \phi_\mathrm{s}(t) \rangle^{3/4}
    \end{align}
    with $e$ the elementary charge, $m_\mathrm{i}$ the ion mass, $\varepsilon_0$ the vacuum permitivity, and $\langle j_\mathrm{i}(t) \rangle$ the averaged ion current inside the sheath.
    For argon, the increased bimodal peak's width is found in figure \ref{fig:IEDF_VolVar90} (left) and for xenon in figure \ref{fig:IEDF_VolVar10} (right).
    From top ($V_\mathrm{RF} = 100\,$V) to bottom ($V_\mathrm{RF} = 1000\,$), the width of the highest energetic bimodal peak increases (Ar\textsuperscript{+} ions: fig. \ref{fig:IEDF_VolVar90} left, Xe\textsuperscript{+} ions: fig. \ref{fig:IEDF_VolVar10} right).\par
    At the same time, a higher driving voltage at a constant pressure produces higher energetic ions.
    Higher kinetic energy enlarges the mean free path of those ions since the mean free path is energy-dependent \cite{ChabertBook} and the collision cross sections fall at high energies (see fig. \ref{fig:crosssections}).
    Thus, the distance between the peaks in the low energetic part of the IEDFs that are connected to charge exchange collisions is increased with the driving voltage \cite{IEDF2, Wild2}.\par
    Furthermore, a raised driving voltage causes the emergence of multiple bimodal structures within the IEDFs.
    These structures have first been reported as double peaks by Wild et al.\cite{Wild2}.
    For our scenario, they become visible for $V_\mathrm{RF} = 500\,$V and $V_\mathrm{RF} = 1000\,$V and establish both for Ar\textsuperscript{+} (fig. \ref{fig:IEDF_VolVar10}) and Xe\textsuperscript{+} ions (fig. \ref{fig:IEDF_VolVar90}).
    Here, charge exchange collisions are responsible for the appearance of the low energetic peak.\par
    Section \ref{GasVar} discusses that low energetic peaks vanish for Ar\textsuperscript{+} ions when the xenon fraction $x_\mathrm{Xe}$ is raised and vice versa.
    For a second or third bimodal peak to establish, two requirements have to be met.
    First, ions have to be able to react to the sheath electric field.
    Second, there has to be some sort of hybrid regime that we discussed in section \ref{GasVar}.
    Combining these requirements also means that only charge exchange collisions that happen clearly above the averaged sheath position can establish an additional bimodal structure.
    Under these conditions, the slow ions produced through charge exchange experience the sheath's modulation that eventually determines their impingement energy.
    A charge exchange collision inside the sheath during the collapsing phase causes the ions to gain slightly lower impingement energy than a charge exchange during the expanding sheath phase. \par
    According to Lieberman and Lichtenberg \cite{LiebermanBook}, there is a weak dependency between the average position of the sheath edge and the voltage amplitude ($s_\mathrm{m} \propto\  V_\mathrm{RF}^{1/4}$).
    Thus, it is more likely for the collisional structures of IEDFs at higher voltages to show bimodal structures.
    The IEDFs of Xe\textsuperscript{+} ions at a xenon fraction $x_\mathrm{Xe} = 0.9$ (fig. \ref{fig:IEDF_VolVar90}) are the best example for this conclusion of our study.
    For $V_\mathrm{RF} = 100\,$V, the results clearly show a single bimodal peak and several non-bimodal charge exchange peaks.
    For $V_\mathrm{RF} = 250\,$V, the main bimodal peak is centered around $\approx 110\,$eV, and at least one additional bimodal peak around $87\,$eV is visible.
    At $500\,$V, the IEDF has at least four bimodal peaks (centered around $\approx 130\,$eV, $\approx 170\,$eV, $\approx 180\,$eV, and $\approx 210\,$eV).
    The case for $V_\mathrm{RF} = 1000\,$V shows at least four bimodal peaks as well (centered around $\approx 190\,$eV, $\approx 260\,$eV, $\approx 325\,$eV, and $\approx 410\,$eV).
    For that case, solely charge exchange collisions that take place deep inside the boundary sheath and close to the electrode do not show any sign of bimodal features.\par
    The hybrid regime of the IEDFs itself is also influenced by a raised driving voltage $V_\mathrm{RF}$.
    For a voltage amplitude $V_\mathrm{RF} = 250\,$V, a slightly higher voltage than that of the base case, the hybrid regime appears for lower admixtures of xenon (fig. \ref{fig:IEDF_VolVar10}) or argon, respectively (fig. \ref{fig:IEDF_VolVar90}).
    Here, the broadening and amplification effects of a raised driving voltage prevail.
    Thus, the hybrid regime establishes earlier than for lower voltages.\par
    The IEDFs for even higher driving voltages (see $V_\mathrm{RF} = 500\,$V and $V_\mathrm{RF} = 1000\,$V in fig. \ref{fig:IEDF_VolVar10} and \ref{fig:IEDF_VolVar90}) are again more collision-dominated and show a different trend.
    For $500\,$V, the bimodal part of the distribution function is less populated than the low energetic part.
    For $1000\,$V, the highest energetic peak for both Ar\textsuperscript{+} (fig. \ref{fig:IEDF_VolVar10}) and Xe\textsuperscript{+} ions (fig. \ref{fig:IEDF_VolVar90}) are damped compared to the lowest energetic peaks.
    This trend arises from the fact that the cross section for charge exchange collisions for high energies drops much slower than the cross sections for isotropic scattering (comp. fig. \ref{fig:crosssections} c) and d)).
    Therefore, charge exchange is the preferred process at high energies.
    For driving voltages much higher than $100\,$V, the hybrid regime is shifted back to higher mixing ratios.

%%%%%%%
\section{Conclusion} \label{conclusion}
The objective of this work was to investigate the ion dynamics of plasmas containing two ion species.
This investigation was conducted by simulating a low-pressure capacitively coupled plasma with a mixture of argon and xenon as the background gas.
The overall result is that the gas composition serves as a means to control the collisionality of the ion species and thus the ion dynamics.
Section \ref{GasVar} shows that the gas composition (more specifically the argon fraction $x_\mathrm{Ar}$ or xenon fraction $x_\mathrm{Xe}$, respectively) significantly affects the discharge, especially the ion dynamics.
The effect on the discharge resembles a parabolic function of the plasma density and the xenon fraction $x_\mathrm{Xe}$ (comp. fig. \ref{fig:ni_GasVar}).
A complete energy balance that we self-consistently calculate based on a PIC/MCC simulation helps understand this effect.
Inelastic processes in xenon (e.g., ionization with $\varepsilon_\mathrm{i,Xe} = 12.12\,$eV) have significantly lower energetic thresholds.
Thus, electrons distribute their energy more efficiently when the xenon fraction $x_\mathrm{Xe}$ is raised.
We show that especially the ionization process in xenon is energetically more favorable than in argon.
This disparity leads to Xe\textsuperscript{+} ions being the dominant ion species for a broad range of xenon fractions $x_\mathrm{Xe}$.\par
For the ion dynamics, we present that the gas composition controls the collisional characteristics of the IEDF.
Between argon and xenon, only non-resonant charge transfer collisions are possible.
Three-body collisions do not occur in relevant amounts in the low-pressure regime.
Therefore, a varied xenon fraction $x_\mathrm{Xe}$ shifts the multiple low energetic peaks (characteristic for charge exchange and a collision dominated regime) from argon (most pronounced at $x_\mathrm{Xe} = 0$, fig. \ref{fig:IEDF_GasVar} left) to xenon (most pronounced at $x_\mathrm{Xe} = 1$, fig. \ref{fig:IEDF_GasVar} right).
Additionally, a collisional/collisionless hybrid regime is present for specific gas fractions.
Some ions experience the discharge within this hybrid regime as collision dominated while others traverse the boundary sheath without collisions.
The analysis of the energy balance helps to understand these effects as well.
It reveals that charge exchange is, even at low-pressures, a relevant energy loss process for ions.
A raised xenon fraction $x_\mathrm{Xe}$ depletes (Ar\textsuperscript{+} ions) or contributes to (Xe\textsuperscript{+} ions) this process for the respective ions (fig. \ref{fig:EB_GasVar}).
Thus, the addition of xenon increases (Ar\textsuperscript{+} ions) or decreases (Xe\textsuperscript{+} ions) the impingement energies of the respective ions.
Furthermore, the energy balance reveals optimal parameters for the impingement energy of ions in this mixture.
In this context, optimal refers to overall minimal collisional losses for the ions, thus desirable conditions for processes (e.g., ion-assisted etching).
For $x_\mathrm{Xe} = 0.4$, the combined fraction of the total energy that ions lose at the surface is maximal.
This example shows that the gas compositions allow tailoring the discharge to the requirements of specific applications.\par
A variation of the driving voltage $V_\mathrm{RF}$ attenuates the dominance of the Xe\textsuperscript{+} ions (sec. \ref{VolVar}).
The reason for this observation is a synergy effect.
The argon's ionization process benefits from additional electrons created during the ionization of xenon.
An extensive analysis of the energy balance is needed to understand this synergy effect and differentiate why it only occurs for Ar\textsuperscript{+} but not Xe\textsuperscript{+}.
Furthermore, it is presented that the increased driving voltage $V_\mathrm{RF}$ intensifies structures (e.g., broadens the width of bimodal peaks) and further complicates the IEDFs (e.g., by creating multiple bimodal peaks).
The energy dependence of the cross section for charge exchange causes the hybrid regime to shift to different mixing ratios when solely varying the driving voltage.
Both observations are supported by the analysis of the energy balance too.
Overall, the energy balance has proven to be a practical and impactful diagnostic.
The results of section \ref{VolVar} show that the gas composition controls the ion dynamics over a wide range of driving voltages.
However, the effect of varied gas compositions is not entirely independent of the driving voltage.\par
Future work based on this study will develop in two directions.
On the one hand, the model system  Ar/Xe needs to be left behind.
The presented basic principles have to be investigated in more complex and process relevant gas mixtures like Ar/CF\textsubscript{4} or CF\textsubscript{4}/H\textsubscript{2}.
The energy balance model can be adapted to and should be tested for these gas mixtures.
On the other hand, based on this work's findings, the influence of a combination of multi-frequency discharges and a varied gas composition on the ion dynamics should be investigated.
For example, a multi-frequency approach could be used to optimize the ion production, which at $V_\mathrm{RF} = 100\,$V was found to be optimal for a xenon fraction $x_\mathrm{Xe} = 0.4$, further.\par
Another open research question is: How does the addition of secondary electron emission and realistic surface coefficients alter the ion dynamics?
The argon ionization's synergy effect, especially, could be significantly affected when secondary electrons cause an amplification of the ionization process.
To our best knowledge, there are no published experimental results that analyze the influence of the gas mixture on the IEDFs.
Nor are there studies that experimentally report about the hybrid regime or the synergy effect within the ionization of argon.
All of these studies would be crucial to validate our findings and simulation.

\ack This work was supported by the German Research Foundation (DFG) via Collaborative Research Centre CRC 1316 (Project ID: 327886311), Transregio TRR87 (Project-ID: 138690629), and project MU 2332/6-1.

\section*{ORCID IDs}
\noindent M. Klich: \href{https://orcid.org/0000-0002-3913-1783}{https://orcid.org/0000-0002-3913-1783}\\
\noindent S. Wilczek: \href{https://orcid.org/0000-0003-0583-4613}{https://orcid.org/0000-0003-0583-4613}\\
\noindent T. Mussenbrock:  \href{http://orcid.org/0000-0001-6445-4990}{http://orcid.org/0000-0001-6445-4990}\\
\noindent J. Trieschmann: \href{http://orcid.org/0000-0001-9136-8019}{http://orcid.org/0000-0001-9136-8019}

%%%%%%%

\end{document}